\documentclass[11pt]{article}
\usepackage[hmargin=1.4cm,vmargin=1.5cm]{geometry}                % Yee geometry.pdf to learn the layout options. There are lots.
\usepackage{graphicx}
\usepackage{amssymb}
\usepackage{amsmath}
\usepackage{amsthm}
\usepackage{epstopdf}
\usepackage{url}
\usepackage{color}
\usepackage{subfigure}
\usepackage{lscape}
\usepackage{authblk}

\newtheorem{proposition}{ Proposition}[section]

\newtheorem{remark}{ Remark}

\newtheorem{definition}{Definition}[section]

\begin{document}

\title{Implied Filtering Densities on Volatility's Hidden State\footnote{This research has been partially supported by NSF grant \#DMS-0739195.}}
%\author[1]{Carlos Fuertes}
%\author[2]{  Andrew Papanicolaou}
% \affil[1]{Bendheim Center for Finance, Princeton University, Princeton NJ, 08544, cfuertes@princeton.edu}
%\affil[2]{Department of Operations Research and Financial Engineering, Princeton University, Princeton NJ 08544, apapanic@princeton.edu}
\author{Carlos Fuertes\footnote{Bendheim Center for Finance, Princeton University, Princeton NJ, 08544, cfuertes@princeton.edu} and Andrew Papanicolaou\footnote{Department of Operations Research and Financial Engineering, Princeton University, Princeton NJ 08544, apapanic@princeton.edu}} 

\date{\today}

\maketitle
\bibliographystyle{plain} 

\maketitle

\begin{abstract}
We formulate and analyze an inverse problem using derivatives prices to obtain an implied filtering density on volatility's hidden state. Stochastic volatility is the unobserved state in a hidden Markov model (HMM) and can be tracked using Bayesian filtering. However, derivative data can be considered as conditional expectations that are already observed in the market, and which can be used as input to an inverse problem whose solution is an implied conditional density on volatility. Our analysis relies on a specification of the martingale change of measure, which we refer to as \textit{separability}. This specification has a multiplicative component that behaves like a risk premium on volatility uncertainty in the market. When applied to SPX options data, the estimated model and implied densities produce variance-swap rates that are consistent with the VIX volatility index. The implied densities are relatively stable over time and pick up some of the monthly effects that occur due to the options' expiration, indicating that the volatility-uncertainty premium could experience cyclic effects due to the maturity date of the options.

\end{abstract}

%\begin{keywords} filtering, stochastic volatility, HMM, volatility risk premium, Heston model, volatility uncertainty.
%\end{keywords}

\bibliographystyle{plain} 
%\bibliographystyle{rAMF} 

%\begin{abstract}

%\end{abstract}
%\tableofcontents

\section{Introduction}
\label{sec:intro}

On the probability space $(\Omega,\mathcal F,\mathbb P)$, define a market consisting of a risk-free asset with constant interest rate $r\geq 0$, and a single stock whose price $S_t$ satisfies the stochastic differential equation
\begin{equation}
\label{eq:dS_physical}
\frac{dS_t}{S_t}=\mu dt+\sigma_tdW_t\ ,
\end{equation}
where $\mu$ is the ex-dividend expected rate of return, $W_t$ is Brownian motion, and $\sigma_t$ is a Markovian stochastic volatility process. Statistical analysis of the observed time series of $S_t$ from equation \eqref{eq:dS_physical} involves a likelihood function that is derived from probability measure $\mathbb P$ that is the \textit{statistical} or \textit{physical} measure. 

The filtration of information in this market is generated by observations of $S$, but realistically we can only assume that the observations occur intermittently at discrete times. The observation times are denoted $\{t_n\}_{n=0,1,2,\dots}$, and we assume (for simplicity) that they are separated by a constant time-step $\Delta t>0$, so that the $n^{th}$ observation time is constructed as $t_n=n\Delta t$. When appropriate, we denote the stock prices and volatility values as 
\begin{align*}
S_n=S_{t_n}\ ,\\ 
\sigma_n=\sigma_{t_n}\ .
\end{align*}
At a time $t\geq 0$, the filtration of market information is denoted $\mathcal F_t^{\Delta t}$ and is formally defined as the $\sigma$-algebra generated by $\{S_n:t_n\leq t\}$. We also write $\mathcal F_n^{\Delta t}$ instead of $\mathcal F_{t_n}^{\Delta t}$ when appropriate. For the class of models that we consider, $\sigma_n$ is not $\mathcal F_n^{\Delta t}$-measurable, and hence not observable, but $(S_n,\sigma_n)$ is a hidden Markov model (HMM) with $\sigma_n$ being the \textit{hidden state}. Given $\mathcal F_n^{\Delta t}$, the posterior distribution of $\sigma_n$ is assumed to have a density 
\[\pi_n(x)\doteq\frac{d}{dx}\mathbb P(\sigma_n\leq x|\mathcal F_n^{\Delta t})\ .\]
In the sequel, we show that the Heston model is an example where the filter indeed has a density (see Proposition \ref{prop:SVMfilter}). The filter is useful for computing estimators of the hidden state, such as the posterior mean (i.e. the optimal posterior estimate in the sense of mean-square error):
\[\hat\sigma_n \doteq \int x\pi_n(x)dx=\arg\min_{\varphi\in\mathcal F_n^{\Delta t}}\mathbb E\left(X_n-\varphi\right)^2\ .\]
A Bayesian method for computing $\pi_n$ can be written down and implemented, but derivative prices are available and are risk-neutral expectations of discounted payoffs given $\mathcal F_n^{\Delta t}$. For instance, a call option is given by% \textcolor{red}{mention equivalent martingale measure}
\[C_n(K) = e^{-r(T-t_n)}\mathbb E^Q[(S_T-K)^+|\mathcal F_n^{\Delta t}]\qquad\hbox{for }t_n\leq T\ ,\]
where $\mathbb E^Q$ is the expectation operator under a risk-neutral probability measure $\mathbb Q$. Thus, a posterior distribution on $\sigma_n$ has already been computed by the market and is embedded in derivatives whose price depends on volatility. Therefore, rather than implement filtering methods on primary market data, we should invert the derivative data to obtain an \textit{implied posterior distribution} on $\sigma_n$. In other words: we solve an inverse problem,
\begin{equation}
\label{eq:introInverseProblem}\min_{\phi\in\mathcal P}\sum_K\left|C_n(K) - e^{-r(T-t_n)}\int \mathbb E^{Q_\sigma}[(S_T-K)^+|S_n,\sigma_n=x]\phi(x)dx\right|^2\ ,
\end{equation}
where $\mathcal P$ is the set of probability densities on $\sigma_n$'s domain and $\mathbb E^{Q_\sigma}$ is the expectation taken under a risk-neutral measure where volatility is observable. This inversion will require some clarification of the risk-neutral measure's structure, which we introduce in Section \ref{sec:sep} as \textit{separability}. We pose the inverse problem as a linear system assuming absence of arbitrage, the Markov property in $(S_t,\sigma_t)$, and the separability condition, but actual computations with data require the additional structure of a stochastic-volatility model as a means for constructing the linear system's model matrix. These matrices %depend on parameters that are also estimated from options data and 
are often of low rank and so the inverse problem must be regularized. Of issue is the bias-variance tradeoff that occurs when solving the inverse problem with regularization, but other issues such as the precision of quoted option prices, uncertainty in the model parameters, and the effects of different degrees of regularization play a role as well. Among our main results is that the inverse problem returns a density with a fair degree of accuracy in the first and second filtering moments, but higher moments are not as accurate.

%\noindent\textbf{Why is it interesting?}\\
The minimizer of equation \eqref{eq:introInverseProblem} is the \textit{implied filtering density} and is denoted $\widehat\phi_n$. It is useful as a proxy for the market's risk-neutral filtering density on volatility, namely, it is a proxy for the posterior density $\frac{d}{dx}\mathbb Q(\sigma_n\leq x|\mathcal F_n^{\Delta t}).$ If we know this risk-neutral filter, then we can proceed to compute other derivative prices, which depend on the hidden state of stochastic volatility. Furthermore, a distribution on $\sigma_n$ with non-zero variance provides a better understanding of volatility uncertainty, which can be helpful in understanding the implied volatility smile. For instance, an implied distribution $\widehat \phi_n$ with non-zero variance contributes additional convexity to the implied volatility smile, a phenomenon that is similar to the result derived in \cite{renaultTouzi1996}. Even though the volatility can be estimated from primary market data, it may be more informative if we look at an implied filtering density that has considered implied volatility data.

\subsection{Relevant Literature}

The relevant background literature for this paper includes many of the statistical studies of stochastic-volatility models, such as the filtering methods in \cite{aiharaBagchi} and \cite{clr2000}, and the parameter estimation methods of \cite{aitSahalia} which also includes careful identification of a parametric mapping between the statistical and risk-neutral models. \cite{christofferson2009} estimate model parameters by calibrating the options data to modeled option prices, and \cite{christofferson2007} estimate parameters and have also implemented a particle filter to obtain a statistical measure on volatility's state; a stochastic volatility particle filter is also used in \cite{viens2003}. With regard to inverse problems, the paper of \cite{carrLee2009} and the paper of \cite{frizGatheral} have identified an inverse problem that is similar to the one we address in this paper, but they have focussed on model-free methods for the case where there is independence between $\sigma_t$ and $W_t^Q$. \cite{cont} have solved a regularized inverse problem to obtain the parameters of a L\'evy measure that drives the underlying. There is also the book by \cite{achdouPironneauBook}, which estimates local volatility surfaces by solving a regularized inverse problem. Relevant empirical works include \cite{figlewski} who has used options data to estimate the risk-neutral density of the underlying, which is also an inverse problem that fits into the framework of this paper. Another is \cite{aitSahaliaMancini} which finds significant jump risk premia in the term structure of variance-swap rates. There is also the paper of \cite{todorov2010}, in which evidence of jumps in realized variance is found. Finally, periodic behavior due to maturity cycles in the implied volatilities of SPX options data is modeled and fit in \cite{FPS2004} with the use of multiple time-scales.

%\noindent\textbf{What have we done?} \\
%\textcolor{red}{In this paragraph you need to spell out what the main results are. What you have found. It is not enough to say "we have explored ...". For example, the sensitivity issue must be spelled out, the role of jumps in %the model, or more generally the role of other models such as multi-term Heston models with or without jumps.} 
\subsection{Preview of Content}

%\textcolor{red}{This section needs a rewrite as the main result of the paper is, in addition to all else, a new and clear interpretation of what this estimated  vol density is .... There should be a clear and precise sentence starting with: "the main result of this paper is" ...\\
%Plus you need to update the section by section content.} 

In this paper, we solve an inverse problem to obtain a non-parametric proxy estimate of volatility's risk-neutral density given the data from the options market. The stochastic-volatility model turns the problem into a linear system, and after some regularization we obtain a solution to equation \eqref{eq:introInverseProblem}, denoted $\widehat \phi_n$, that is the \textit{implied density or implied filter} on volatility. Our assumption of \textit{separability} in the martingale change of measure leads to a convenient framework in which volatility uncertainty can be priced and an uncertainty premium inferred from options data. 

From real data on the SPX index and its options from the year 2005, we find some interesting results when solving the inverse problem with the Heston model. In particular, maturity cycles appear in the risk-neutral standard deviation of $\sigma_n$, indicating a connection between the risk premium on volatility uncertainty and the `little-t' effects observed in \cite{FPS2004}. We also consider a Heston model with jumps and find that neither $\widehat\phi_n$ or the fitted parameters exhibit the maturity cycles in a clear way, which suggests that the volatility-uncertainty premium has perhaps been concealed by an over fitting of the data. %We also compute the expectations of 30-day variance under the implied measure, and find that the inclusion of jumps adds a risk premium over the VIX's price on variance.

% \textcolor{red}{You need to be more precise in what you have found here: For example, that the result in section 5.1 gives only the mean of your conditional probability measure, so the model free theory is not only sensitive to having to differentiate observed option prices with respect to strike time and strike, which is a sensitive interpolation issue, it also does not tell you what the variance of this measure might be, except when $\rho=0$, which is of limited value. So getting estimates of the full measure is relevant. Another thing to point out is behavior with respect to $t$: it is much more complicated than what filtering might give you ... Also, the very interesting connection of Fig 9 with the maturity cycles paper: Maturity cycles in implied volatility. Jean-Pierre Fouque, George Papanicolaou, K. Ronnie Sircar and K. Solna, Finance and Stochastics, 8, (2004), pp. 451-477. Clearly the jumps are trying to take out the cycles (but more work needs to be done here ...) and your approach with the vol measures gives a very natural way to assess the success of the model, namely the fact that the std (bottom right) with the jumps is consistently small over time. This is a "major" result in my view and one for which you have no competition. It makes a very strong case for your approach.}

The rest of the paper is organized as follows: Section \ref{sec:inverseProblem} presents the problem, defines separability of the martingale change of measure, introduces the notation that is used throughout, and provides some insight into the Bayesian nature of the problem; Sections \ref{sec:inverseProblemPosed} and \ref{sec:illPosedness} discuss ill-posedness and the Tykhonov regularization, with Sections \ref{sec:blackScholesExample} and \ref{sec:hestonToyExample} presenting some basic examples with the Black-Scholes and Heston models; Section \ref{sec:error} describes the error that is caused by measurement imprecision in option prices, and the error caused by uncertainty in the stochastic-volatility model's parameters; in Section \ref{sec:data}, a Heston model and a Heston model with jumps are fitted to the SPX data, the inverse problem is solved, a comparison of the results is presented, the periodic behavior among short time-to-maturity options is pointed out, and the fitted model's variance-swap rate is compared to the VIX index. 
%\noindent\textbf{What is the next move?}\\
%\subsection{Possible Research for the Future}
%In the future, more analysis of the inverse problem should be done, both analytically and with simulations. Other applications of the inverse problem should be explored; one example would be commodities where convenience yield could be treated as a hidden Markov state in futures prices. Empirically, more models and more data should be investigated. This paper has only considered the SPX and VIX from 2005, but the same analysis can be performed on the QQQQ and VXN, or the \^{}OEX and VXO, and it would also be interesting to see how the implied measures behaved during The Crisis. In particular, data should be analyzed with an alertness for the maturity cycles, and whether or not they come out as more or less pronounced.
% 

%%%%%%%%%%%%%%%%%%%%%%
\section{Stochastic-Volatility Model \& The Inverse Problem}
\label{sec:inverseProblem}

%\textcolor{red}{The risk neutral system has a form that depends on $\rho$. Look at the FPSS book (Ch 3) for how
%it is written there. Look also at what you have later in Section 4.1}

In this section we add stochastic volatility to the model described in Section \ref{sec:intro}, and also clarify some of the statements that were made regarding the observability of volatility and the existence of a filtering density. The SDEs for $S_t$ are redefined to include specification of the stochastic volatility process, but other elements that were considered in Section \ref{sec:intro} are carried-over, such as the discrete nature of information, the filtration $\mathcal F_n^{\Delta t}$, the filtering density $\pi_n$, and the implied density $\widehat \phi_n$. 

Our stochastic-volatility model involves the jointly-Markov process $(S_t,X_t)$, with $S_t$ denoting the asset's price, and $X_t$ being the state of volatility. The model is
\begin{equation}
\label{eq:svm}
\begin{array}{ccl}
dS_t&=& \mu S_t dt+\sigma(X_t)S_t(\rho dB_t+\sqrt{1-\rho^2}dW_t)\ ,\\
&&\\
dX_t&=&a(X_t)dt+b(X_t)dB_t\ ,
\end{array}
\end{equation}
where $(W_t,B_t)$ are independent Brownian motions, the parameter $\rho$ is the volatility-leverage effect with $\rho\in(-1,1)$, the functions $a$ and $b$ are the drift and volatility-of-volatility, respectively, and $\sigma(x)\in C(\mathbb R^+)$ is a continuous and one-to-one function such that $\sigma(X_t)$ is the volatility at time $t\geq 0$ . The process $(S_t,X_t)$ is measurable on the probability space $(\Omega,\mathcal F,\mathbb P)$, but we assume there is no arbitrage so that there exists an equivalent martingale measure $\mathbb Q\sim \mathbb P$ under which the discounted value of $S_t$ is a martingale (see chapter 10 of \cite{bjork}). The risk-neutral model is
\begin{equation}
\label{eq:svm_riskNeutral}
\begin{array}{ccl}
dS_t &=& rS_t dt+\sigma(X_t)S_t(\rho dB_t^Q+\sqrt{1-\rho^2}dW_t^Q)\ ,\\
&&\\
dX_t&=&a(X_t)dt+\lambda(X_t,S_t,t)dt+b(X_t)dB_t^Q\ ,
\end{array}
\end{equation}
with $(W_t^Q,B_t^Q)$ being independent $\mathbb Q$-Brownian motions, and with $X_t$'s drift being altered by the market's price of volatility risk in $\lambda(x,s,t)$. A popular example is the Heston model where $\sigma(x)=\sqrt x$ and $X_t$ is a Cox-Ingersol-Ross (CIR) process
\begin{align}
\nonumber
dX_t &= \kappa(\bar X-X_t)dt-\lambda X_t+\gamma\sqrt{X_t}dB_t^Q\\
\nonumber
&=(\kappa\bar X-(\kappa+\lambda)X_t)dt+\gamma\sqrt{X_t}dB_t^Q\\
\label{eq:hestonVolRiskNeutral}
&=(\kappa+\lambda)\left(\frac{\kappa\bar X}{\kappa+\lambda}-X_t\right)dt+\gamma\sqrt{X_t}dB_t^Q\ ,
\end{align}
where $\kappa>0$, $\bar X>0$ and $\gamma>0$, and historical estimates of the price of volatility risk in $\lambda$ are often found to be negative, indicating that the risk-neutral average of $\sigma(X_t)$ is higher than the statistical average. An important condition in the Heston model is the Feller condition $\gamma^2\leq 2\bar X\kappa$, which ensures that $X_t>0$ for all $t>0$ with probability 1 under the $\mathbb P$-meausure (see chapter 3 of \cite{FPS2011}), and also insures $X_t>0$ for all $t>0$ with probability 1 under the $\mathbb Q$-measure provided that $\kappa+\lambda>0$.

%\textcolor{red}{Connect the above to what you have later in Section 4.1}

The majority of results on stochastic volatility have been obtained in continuous time, where volatility is observed and hedging portfolios can be continuously rebalanced. The reason why volatility is observed in continuous time is because the time derivative of $\log(S_t)$'s quadratic variation is observed:

\begin{equation}
\label{eq:observe}
\sigma^2(X_t) =\frac{d}{dt}\left<\log(S)\right>_t\ ,
\end{equation}
where $\left<~\cdot~\right>$ denotes \textit{quadratic variation}, which an observed quantity given the filtration generated by continuous-time observations $\{S_u:u\leq t\}$. The right-hand side of equation \eqref{eq:observe} is a limit in probability of the squared differentials of $\log(S_t)$ and is equal to $\sigma^2(X_t)$ almost-everywhere in time. Prices, however, are quoted discretely by the market which means that the right-hand side of equation \eqref{eq:observe} may at best be estimated from the time series $(S_\ell)_{\ell=0}^n$. Hence, volatility is an unobserved process that needs to be estimated from the information in $\mathcal F_n^{\Delta t}$. Moreover, volatility information from the options market (i.e. the implied volatility information) is absent in estimates that rely on the time series alone. In the coming sections we present a methodology that incorporates the implied information into the estimation procedure.

\subsection{Filtering}
Filtering is a way to track the hidden state of an HMM, and our stochastic-volatility model is in fact an HMM. This can be confirmed through simple inspection of the model whereby one verifies that $(S_n,X_n)\doteq (S_{t_n},X_{t_n})$ is a Markov process. Furthermore, it was mentioned in Section \ref{sec:intro} that filters often have densities, and from the general specification set forth in equation \eqref{eq:svm} we can say that existence of a filtering density ultimately depends on the specification of functions $a(x)$, $b(x)$ and $\sigma(x)$. The following theorem shows that a filtering density exists for the Heston model, and the same proof is applicable to other models:

%%%%%%%% theorem on Heston Vol Filter
\begin{proposition} 
\label{prop:SVMfilter}
Let $(S_t,X_t)$ be the price and volatility processes in the Heston model referred to in equation \eqref{eq:hestonVolRiskNeutral}, and assume the Feller condition $\gamma^2\leq 2\bar X\kappa$. Then there is a kernel $\Gamma$ that gives $X$'s transition density, call it $\Gamma^{\Delta t}(x|v)\doteq\frac{d}{dx}\mathbb P(X_{t+\Delta t}\leq x|X_t=v)$ for any $x,v\in\mathbb R^+$, and the filtering distribution for $X_n$ at observation time $t_n=n\Delta t$ has a density. This density is given recursively as
\begin{align}
\nonumber
&\pi_n(x)\\
\label{eq:piDensity}
&=\frac{1}{c_n}\int \mathbb E\left[\mathbb L(y|(X_u)_{\{t_{n-1}\leq u\leq t_n\}},S_{n-1})\Big|X_n=x,X_{n-1}=v,S_{n-1}\right]\Gamma^{\Delta t}(x|v)\pi_{n-1}(dv)\Bigg|_{y=S_n}\ ,
\end{align}
for almost-everywhere $x\in\mathbb R^+$, where $c_n$ is a normalizing constant, and $\mathbb L$ is the likelihood of the path $(x_u)_{\{t_{n-1}\leq u\leq t_n\}}$ given observations $S_n$ and $S_{n-1}$, and is given by 
\[\mathbb L(y|(x_u)_{\{t_{n-1}\leq u\leq t_n\}},S_{n-1}) =\frac{\exp\left\{ -\frac{1}{2}\left(\frac{ \log(y/S_{n-1})-\left(\mu\Delta t-.5\int_{t_{n-1}}^{t_n}x_udu\right)-\rho\xi_n(x)}{\sqrt{(1-\rho^2)\int_{t_{n-1}}^{t_n}x_udu}}\right)^2 \right\}}{ \sqrt{(1-\rho^2)\int_{t_{n-1}}^{t_n}x_udu}}\]
with 
\[\xi_n(x) = \frac{1}{\gamma}\left\{ \Delta x_{n-1}-\kappa\left(\bar X\Delta t-\int_{t_{n-1}}^{t_n}x_udu\right)\right\}\ .\]
\end{proposition}

\begin{proof} Given the Feller condition, the CIR process $dX_t=\kappa(\bar X-X_t)dt+\gamma\sqrt{X_t}dB_t$ is well-known to have a transition density that can be written in terms of a modified Bessel function (see \cite{aitSahalia1999}), and so $\Gamma^{\Delta t}(\cdot|v)$ is a smooth density function for all $v\geq 0$. Furthermore, it was shown in \cite{DY2002} that $(S_t,X_t)$ has a smooth transition density function, that is, 
\[\mbox P^{\Delta t}(y,x|s,v)\doteq \frac{\partial^2}{\partial y\partial x}\mathbb P(S_n\leq y,X_n\leq x|S_{n-1}=s,X_{n-1}=v)\]
is smooth for $x> 0$, $y> 0$, and $\Delta t>0$, and does not collect mass at $x=0$ or $y=0$. Hence, the filter has a density that can be written using Bayes rule:
\[\pi_n(x) = \frac{\int\mbox P^{\Delta t}(S_n,x|S_{n-1},v)\pi_{n-1}(v)dv}{\int\hbox{[numerator]}dx}\ ,\]
where we don't need to assume smoothness of $\pi_{n-1}$ because it is smoothed against $\mbox P^{\Delta t}$ in the $dv$-integral.

Now, from equation \eqref{eq:svm} we notice the following:
\begin{align*}
\log(S_n/S_{n-1}) &=  \mu\Delta t-\frac 12\int_{t_{n-1}}^{t_n}X_udu+\rho\int_{t_{n-1}}^{t_n}\sqrt{X_u}dB_u+\sqrt{1-\rho^2}\int_{t_{n-1}}^{t_n}\sqrt{X_u}d W_u\\
&=_d  \mu\Delta t-\frac 12\int_{t_{n-1}}^{t_n}X_udu+\rho\int_{t_{n-1}}^{t_n}\sqrt{X_u}dB_u+\sqrt{(1-\rho^2)\int_{t_{n-1}}^{t_n}X_udu}~~\mathcal Z
\end{align*}
where ``$=_d$" signifies equivalence in distribution, and $\mathcal Z$ is another independent standard normal random variable. This means that conditional on the path $(X_u)_{t_{n-1}\leq u\leq t_n}$ and $S_{n-1}$, 

\[\frac{\log(S_n/S_{n-1})-\left(\mu\Delta t-\frac 12\int_{t_{n-1}}^{t_n}X_udu+\rho\int_{t_{n-1}}^{t_n}\sqrt{X_u}dB_u\right)}{\sqrt{(1-\rho^2)\int_{t_{n-1}}^{t_n}X_udu}}=_d\mathcal Z\ .\]
Then noticing $\xi_n$ evaluated at $(X_u)_{t_{n-1}\leq u\leq t_n}$ is the the same as $\xi_n\left(X\right)=\int_{t_{n-1}}^{t_n}\sqrt{X_u}dB_u$, it follows that
\[\frac{\log(S_n/S_{n-1})-\left(\mu\Delta t-\frac 12\int_{t_{n-1}}^{t_n}X_udu+\rho\xi_n\left(X\right)\right)}{\sqrt{(1-\rho^2)\int_{t_{n-1}}^{t_n}X_udu}}=_d\mathcal Z\ .\]
This shows the likelihood of the path $(X_u)_{t_{n-1}\leq u\leq t_n}$ given $S_{n-1}$ and $S_n=y$ is in fact the function $\mathbb L$.

Finally, given Bayes rule for the density $\pi_n$, the expression in equation \eqref{eq:piDensity} displays the filter using a probabilistic representation of the transition density:

\begin{align*}
&\mbox P^{\Delta t}(y,x|S_{n-1},v)\\
&=\frac{\partial}{\partial y}\int_0^y\mbox P^{\Delta t}(z,x|S_{n-1},v)dz \\
& =\frac{\partial}{\partial y}\mathbb P(S_n\leq y|X_n=x,X_{n-1}=v,S_{n-1})\Gamma^{\Delta t}(x|v)\\
& =\frac{\partial}{\partial y}\mathbb E\left\{\mathbf 1_{S_n\leq y}\Big |X_n=x,X_{n-1}=v,S_{n-1}\right\}\Gamma^{\Delta t}(x|v)\\
&=\frac{\partial}{\partial y}\mathbb E\left[\mathbb E\left\{\mathbf 1_{S_n\leq y}\Big|( X_u)_{\{t_{n-1}\leq u\leq t_n\}},S_{n-1}\right\}\Big| X_n=x, X_{n-1}=v,S_n,S_{n-1}\right]\Gamma^{\Delta t}(x|v)\\
&=\mathbb E\left[\frac{\partial}{\partial y}\mathbb E\left\{\mathbf 1_{S_n\leq y}\Big|( X_u)_{\{t_{n-1}\leq u\leq t_n\}},S_{n-1}\right\}\Big| X_n=x, X_{n-1}=v,S_n,S_{n-1}\right]\Gamma^{\Delta t}(x|v)\\
&\propto\mathbb E\left[\mathbb L(y|(X_u)_{\{t_{n-1}\leq u\leq t_n\}},S_{n-1})\Big|X_n=x, X_{n-1}=v,S_{n-1}\right]\Gamma^{\Delta t}(x|v)\ .
\end{align*}
Lastly, when computing the likelihood based on the time-$n$ observation, the last line is evaluated at $y=S_n$. This completes the proof of the proposition.

%and so we must also take independent copies $\tilde X$ to avoid bias in the sampling of the paths $(X_u)_{\{t_{n-1}\leq u\leq t_n\}}$. Hence, $\tilde X$ replaces $X$ in the last line of the display.
% The filtering measure exists and is defined with a Bayesian recursion,
%\begin{align*}
%&\mathbb P\left(X_n\in\mathcal A\Big|\mathcal F_n^{\Delta t}\right)\\
%& = \frac{1}{c_n}\int_{\mathcal A}\int \mathbb E\left[\mathbb P(S_n|(\tilde X_u)_{\{t_{n-1}\leq u\leq t_n\}},S_{n-1})\Big|\tilde X_n=x,\tilde X_{n-1}=v,S_n,S_{n-1}\right]\Gamma^{\Delta t}(x|v)\pi_{n-1}(dv)dx\ ,
%\end{align*}
%for any Borel set $\mathcal A\subset\mathbb R^+$, with $\tilde X$ being an independent copy of $X$, and with $c_n$ being a normalizing constant. One way to show that $\pi_n$ has a density is with an application of the Lebesgue differentiation theorem, which applies here because the integrand inside the $dx$ integral is integrable (i.e. because it is non-negative and integrates over $\mathbb R^+$ to 1). Hence, the Lebesgue differentiation theorem says
%\[\frac{1}{\left|\mathcal A\right|}\mathbb P\left(X_n\in\mathcal A\Big|\mathcal F_n^{\Delta t}\right)\rightarrow \pi_n(x)\]
%as $\mathcal A$ \textit{shrinks nicely} to the singleton set $\{x\}$ for almost-everywhere $x\in\mathbb R^+$ (see \cite{folland} for the Lebesgue theory on integration)
\end{proof}

%%%%%%%%%%%%%%%
\subsection{Separability of the Martingale Change of Measure}
\label{sec:sep}
Given the filtration $\mathcal F_n^{\Delta t}$, the risk-neutral price of a derivative contract is a conditional expectation of the payoff. For instance, a call option with strike $K$ is given by
\[C_n(K) \doteq e^{-r(T-t_n)}\mathbb E^Q\left\{(S_T-K)^+\big|\mathcal F_n^{\Delta t}\right\}\qquad\forall t_n\leq T\ .\]
If $\sigma_n$ were $\mathcal F_n^{\Delta t}$-measurable, then the Markov property would apply and the entire history of observations would not be necessary; we could simply write the price as an expectation conditional on $S_n$ and $\sigma_n$. Since $S_n$ is not Markovian by itself, and it is all that is observed, we therefore must write $\mathcal F_n^{\Delta t}$ in the conditioning for the risk-neutral expectation.

Since $S_T$ is observable, suppose that $T=N\Delta t$, for some integer $N$, so that $S_N=S_{t_{N}}=S_T$.  The equivalent martingale change of measure is defined by an $\mathcal F_n^{\Delta t}$-adapted martingale $Z_n$, such that
\begin{equation}
\label{eq:nonMarkovPrice}
\mathbb E^Q\left\{(S_{N}-K)^+\big|\mathcal F_n^{\Delta t}\right\} =\mathbb E\left\{\frac{Z_{N}}{Z_n}(S_{N}-K)^+\Big|\mathcal F_n^{\Delta t}\right\}\ .
\end{equation}
Equation \eqref{eq:nonMarkovPrice} is the main premise for the work in this paper because it shows how risk-neutral pricing is a filtering expectation. However, incompleteness of the market means that the martingale $Z_n$ is non-unique and so the market needs to decide on the price of volatility risk. The price of volatility risk is standard in the stochastic volatility literature, but should also include an additional premium if there is volatility uncertainty (i.e. if the volatility process is unobserved). In the sequel it is important that we are able to separate the premium on volatility uncertainty from $Z_n$, which we assume to be the case, in order to solve the inverse problem and obtain the implied filtering density.

Let $\mathcal G_n $ denote a larger filtration such that $\mathcal F_n^{\Delta t}\subset \mathcal G_n$ under which $\sigma_n$ is $\mathcal G_n$-measurable. The assumption required to solve the inverse problem is a condition in the martingale change of measure, which we call \textit{separability}:
%%%%%%%%%%% separable martingale
\begin{definition} 
\label{def:separable}\textbf{(Separability)}
The martingale change of measure $Z_n$ that is adapted to the filtration $\mathcal F_n^{\Delta t}$, is considered separable if it can be specified as the product of an $\mathcal F_n^{\Delta t}$-adapted Radon-Nykodym derivative and another $\mathcal G_n$-adapted martingale change of measure. We write this as
 \[\frac{Z_{N}}{Z_n}=\Lambda_n(X_n)\frac{\mathcal M_{N}}{\mathcal M_n}\ ,\]
 where $\Lambda_n(x)$ is $\mathcal F_n^{\Delta t}$-adapted for a.e. $x$ for all $n\geq 0$, and $\mathcal M_n$ is a martingale change of measure with $\mathbb E\left[\mathcal M_{n+1}\Big|\mathcal G_n\right] =\mathcal M_n$ for all $n\geq 0$.

\end{definition}
\begin{remark}
\label{rem:reconcile}
Let $\mathcal F_t $ denote the filtration generated by $\{S_u:u\leq t\}$, under which $X_t$ is observable. In the case where $\mathcal G_n=\mathcal F_{t_n}$, Definition \ref{def:separable} is a way of reconciling the continuous time theory on option pricing with the reality that trading cannot occur in continuos time (i.e. continuous time hedging portfolios cannot be perfectly maintained) and that volatility is not observable. This idea is of fundamental importance for the rest of this paper: our numerical experiments and data analysis assume separability of $Z_n$ with an equivalent martingale measure associated with the continuous time stochastic-volatility model presented in equation \eqref{eq:svm}.
\end{remark}

\begin{remark}
\label{rem:riskPrem}
Our interpretation is that $\Lambda_n$ is a risk premium on volatility uncertainty. Indeed, this viewpoint is not contradicted by the experiments of Sections \ref{sec:blackScholesExample}, \ref{sec:hestonToyExample}, and in the data analysis of Section \ref{sec:data}. Our experiments rely on the Heston model and assume separability of $Z_n$ with the Heston model's martingale change of measure. In particular, the data from European options on the S\&P500 implies periodic behavior in the time series of moments of the risk-neutral filter. This periodicity is related to the monthly maturity cycles of the options, and we suspect it to be the implied $\Lambda_n$ that accounts for a possible increase in premium at times near maturity. More is said about maturity effects in Section \ref{sec:matCycles}.%We suspect that $\Lambda_n$ is the cause of this periodic behaviour because maturity effects cannot be present in $\pi_n$, as $\pi_n$ a physical measure that is obtained from time series of stock prices that has no maturities. 
\end{remark}

\begin{remark}
\label{rem:misSpecified}
Among other things, Definition \ref{def:separable} is a specification of the martingale change of measure defining $\mathbb Q$. It is possible that market data will reject the hypothesis that Definition \ref{def:separable} holds, in which case any model assuming separability of $Z_n$ is \textit{mis-specified}. This paper does not do any testing of the hypothesis, but certainly it should be something that the reader is aware of --particularly when reading Section \ref{sec:data} where we look at real market data.
\end{remark}
As alluded to in Remark \ref{rem:reconcile}, the condition of separability set forth in Definition \ref{def:separable} is useful because it means that the option price can be written as an average of a continuous time model's option price. This is demonstrated in the following proposition: 

\begin{proposition}
\label{prop:iteratedExpectations}
Consider the filtration $\mathcal F_t $ generated by $\{S_u:u\leq t\}$, and let $\mathcal M_t$ be an equivalent martingale measure adapted to $\mathcal F_t$. Suppose $Z_n$ satisfies Definition \ref{def:separable} with respect $\mathcal F_n$ and the $\mathcal F_n$-adapted martingale $\mathcal M_n=\mathcal M_{t_n}$, and define $\phi_n$ to be 
\begin{equation}
\label{eq:phi_n}
\phi_n(x) \doteq \Lambda_n(x)\pi_n(x)
\end{equation}
for almost-everywhere $x$ in $\sigma_n$'s domain. Then the price of a European call option can be written in terms of iterated expectations:
\begin{equation}
\label{eq:iteratedExpectations}
C_n(K) = e^{-r(T-t)}\int_{\mathbb R^+}\mathbb E^{Q_{\sigma}}\left[(S_{N}-K)^+\big|S_n,X_n=x\right]\phi_n(x)dx\ ,
\end{equation}
where the inner expectation is taken under an equivalent martingale measure $\mathbb {Q_{\sigma}}$ that prices in the continuous-time setting where $\sigma_n$ is observed.
\end{proposition}
\begin{proof}
\begin{eqnarray}
\nonumber
C_n(K)& =& e^{-r(T-t)}\mathbb E^Q\left[ (S_{N}-K)^+\Big|\mathcal F_n^{\Delta t}\right]\\
\nonumber
&=& e^{-r(T-t)}\mathbb E\left[\frac{Z_{N}}{Z_n}(S_{N}-K)^+\Big|\mathcal F_n^{\Delta t}\right]\\
\nonumber
&=& e^{-r(T-t)}\mathbb E\left[\mathbb E\left[\frac{Z_{N}}{Z_n}(S_T-K)^+\Big|\mathcal F_n\right]\Big|\mathcal F_n^{\Delta t}\right] \qquad\hbox{because }\mathcal F_n^{\Delta t}\subset\mathcal F_n,\\
\nonumber
&=& e^{-r(T-t)}\mathbb E\left[ \mathbb E\left[\Lambda_n(X_n)\frac{\mathcal M_{N}}{\mathcal M_n}(S_{N}-K)^+\Big|\mathcal F_n\right]\Big|\mathcal F_n^{\Delta t}\right]\\
\nonumber
&=& e^{-r(T-t)}\mathbb E\left[\Lambda_n(X_n) \mathbb E\left[\frac{\mathcal M_{N}}{\mathcal M_n}(S_{N}-K)^+\Big|\mathcal F_n\right]\Big|\mathcal F_n^{\Delta t}\right]\\
\nonumber
&=& e^{-r(T-t)}\mathbb E\left[\Lambda_n(X_n) \mathbb E\left[\frac{\mathcal M_{N}}{\mathcal M_n}(S_{N}-K)^+\Big|S_n,X_n\right]\Big|\mathcal F_n^{\Delta t}\right]\qquad\hbox{by the Markov property,}\\
\nonumber
&=& e^{-r(T-t)}\int\mathbb E\left[\frac{\mathcal M_{N}}{\mathcal M_n}(S_{N}-K)^+\Big|S_n,X_n=x\right]\Lambda_n(x)\pi_n(x)dx\\
\nonumber
&=& e^{-r(T-t)}\int\mathbb E^{Q_{\sigma}}\left[(S_{N}-K)^+\big|S_n,X_n=x\right]\phi_n(x)dx\ ,
\end{eqnarray}
where $\mathbb E^{Q_\sigma}$ is the expectation operator under the equivalent martingale measure $\mathbb Q_\sigma$ that is defined by $\mathcal M_t$.
\end{proof}
Given options data from the market, option pricing formulae from continuous time (observable) stochastic-volatility models can be used to invert equation \eqref{eq:iteratedExpectations} in order to obtain an estimate of $\phi_n$ defined by equation \eqref{eq:phi_n}. This inverse problem is formulated in the next section, and assumes a priori the separability condition of Definition \ref{def:separable}. If separability does not hold, then the method presented here does not apply, but statistical tests for rejection of the hypothesis  ``$Z_n$ is separable" is an interesting problem in itself (see Remark \ref{rem:misSpecified}).

Finally, it should also be pointed out that $\phi_n = \mathbb Q(~\cdot~|\mathcal F_n^{\Delta t})$ if $\mathbb E^{Q_{\sigma}}\left[(S_{N}-K)^+\big|S_n,\sigma_n=x\right] = \mathbb E^{Q}\left[(S_{N}-K)^+\big|S_n,\sigma_n=x\right]$ for almost-everywhere $x$ and pointwise in $S_n$. In other words, if $\mathbb Q_\sigma = \mathbb Q$, then $\phi_n$ is the risk-neutral filter. The numerical experiments of Sections \ref{sec:blackScholesExample}, Section \ref{sec:hestonToyExample} and Section \ref{sec:error} assume that $\mathbb Q_\sigma = \mathbb Q$, but such an assumption cannot be made in Section \ref{sec:data} where we work with real data. In Section \ref{sec:data}, $\widehat\phi_n$ is not necessarily an estimator of the risk-neutral filter, but nonetheless we use it as a \textit{proxy} for $\mathbb Q(~\cdot~|\mathcal F_n^{\Delta t})$.
%\textcolor{red}{Give a number to this formula (rewrite first-left and last) and integrate it into the paper as the main result that motivates the inverse problem. This tells you what exactly it is that you are calculating!\\
%Also, you should use the notation $\mathbb {Q_{\sigma}}$ consistently later when it comes up. You use it OK in the
%next section.}

%%%%%%%%%%%%%%%%%%%%%%%%%%%%

\subsection{Inverse Problem}
\label{sec:inverseProblemPosed}

Our approach to the inverse problem uses a stochastic-volatility model, but the specification of this model is left open. The only assumptions are (i) that $(S_n,X_n)$ is a Markov process, and (ii) that the equivalent martingale measure $\mathbb Q$ satisfies the separability condition of Definition \ref{def:separable}. Standard results on stochastic volatility have derived formulae for European call options as a function of $(S_n,X_n)$:

\[C(K,T,t_n,S_n,X_n) \doteq e^{-r(T-t_n)}\mathbb E^{Q_x}\left[(S_T-K)^+\Big|S_n,X_n\right]\qquad\qquad\hbox{for all }t_n\leq T\ ,\]
where $\mathbb E^{Q_x}$ denotes the risk-neutral expectation in a setting where $X$ is observed. Based on Proposition \ref{prop:iteratedExpectations}, equation \eqref{eq:iteratedExpectations} can be written as an average of the function $C(K,T,t_n,S_n,~\cdot~)$:
\begin{equation}
\label{eq:hullWhite}
C_n(K,T)\doteq e^{-r(T-t_n)}\mathbb E^Q\left[(S_T-K)^+\Big|\mathcal F_n^{\Delta t}\right] = \int C(K,T,t_n,S_n,x)\phi_n(dx)\qquad\qquad\hbox{for all }t_n\leq T.
\end{equation}
Our convention is to identify the call options with the indices $i=1,2,\dots,M$ so that the strike and maturity of the $i^{th}$ option are $K_i$ and $T_i$, respectively. Then, based on equation \eqref{eq:hullWhite} we invert the set of call options to obtain $X_n$'s implied filtering density, 
\begin{equation}
\label{eq:inverseProblem}
\widehat\phi_n\doteq\arg\min_{\phi\in\mathcal P}\sum_{i=1}^M\left|C_n(K_i,T_i)-\int C(K_i,T_i,t_n,S_n,x)\phi(x)dx\right|^2\ ,
\end{equation}
where $\mathcal P$ is the set of probability densities on $X_n$'s domain.

In general, we can set up the inverse problem in equation \eqref{eq:inverseProblem} using any kind of derivative on $S$ provided that the pricing formula fits the framework for equation \eqref{eq:hullWhite}. For instance, we could also include put options, variance swaps, futures, Asian options, or any other European claim. We could also set up the inverse problem as a search across a family of parametric distributions (e.g. $\mathcal P$ is some exponential family), but this paper is a study of the inverse problem where the search looks for a non-parametric estimate of the probability measure.
%\[\mathcal P=\left\{\phi:\phi(x)\geq 0,~\forall x\in\mathbb R^+~\hbox{and}~\int \phi(x)dx=1\right\}.\]

\begin{remark} The solution to the inverse problem is a measure that is \textit{implied} by the options market. Much like implied volatility in the Black-Scholes model, the solution to equation \eqref{eq:inverseProblem} is an estimate of the input to a formula for \textit{modeled} market prices. We fully expect solutions of the inverse problem to contain evidence of the model's shortcomings when real market data is used as input. Much like Black-Scholes implied volatility, we need to find a way to interpret this evidence in terms of market effects that have not been modeled.
\end{remark}

%%%%%%%%%%%%%%%
\subsection{Ill-Posedness}
\label{sec:illPosedness}
Consider the following notation for the option prices at observation time $t_n$:
\[C_n^i\doteq C_n(K_i,T_i) ~~~\hbox{and}~~~ C^i(t_n,S_n,x) \doteq C(K_i,T_i,t_n,S_n,x).\] 
Numerically, solving the inverse problem in equation \eqref{eq:inverseProblem} amounts to a finite linear system whose solution is a set of discrete weights on various sample points in $X_n$'s domain. We discretize by choosing a set of sample-points $x_1<x_2<~\dots~<x_{\mathcal H}$ (where $\mathcal H\in\mathbb Z^+$ is the integer-size of $x$'s numerical domain), and solve the following system to obtain the posterior distribution,
\begin{equation}
\label{eq:linSystemScalar}
\begin{bmatrix}
C_n^1\\
C_n^2\\
%C_n^3\\
\vdots\\
\vdots\\
C_n^M
\end{bmatrix}=
\begin{array}{cc}
\begin{bmatrix}
C^1(t_n,S_n,x_1)&C^1(t_n,S_n,x_2)&\dots&\dots&C^1(t_n,S_n,x_{\mathcal H})\\
C^2(t_n,S_n,x_1)&C^2(t_n,S_n,x_2)&\dots&\dots&C^2(t_n,S_n,x_{\mathcal H})\\
%C^3(S_t,x_1,t)&C^3(S_t,x_2,t)&\dots&\dots&C^3(S_t,x_{\mathcal H},t)\\
\vdots&\vdots&\ddots&&\vdots\\
\vdots&\vdots&&\ddots&\vdots\\
C^M(t_n,S_n,x_1)&C^M(t_n,S_n,x_2)&\dots&\dots&C^M(t_n,S_n,x_{\mathcal H})
\end{bmatrix}\cdot
\begin{bmatrix}
\phi_n(n_{x_1})\\
\phi_n(n_{x_2})\\
%\phi_n^Q(x_3)\Delta x_3\\
\vdots\\
\vdots\\
\phi_n(n_{x_{\mathcal H}})\\
\end{bmatrix}
\end{array}+\epsilon\ ,
\end{equation}
where $\{n_{x_j}\}_j$ denotes a set of disjoint neighborhoods with $x_j\in n_{x_j}$ $\forall j$ such that $\phi_n(n_{x_j}) \geq 0$, $\phi_n(n_{x_i}\cap n_{x_j})=0$ for $i\neq j$, and $\sum_j\phi_n(n_{x_j})=1$. The quantity $\epsilon\in\mathbb R^M$ is a small noise vector that is orthogonal to the columns of the design matrix and is attributed to various inconsistencies such as
\begin{itemize}
\item measurement imprecision, which occurs because prices are quoted only to 2 decimal places,
\item parameter estimation error (e.g. coefficients of the SDE for stochastic volatility are unknown),
\item and numerical integration error caused by discretization.
\end{itemize}
A detailed discussion of error is the topic of Section \ref{sec:error}. For now we assume that $\epsilon$ is an idiosyncratic component that does not affect optimization. In matrix/vector notation, equation \eqref{eq:linSystemScalar} can be written as
\begin{equation}
\label{eq:linSystem}
C_n = C\phi_n+\epsilon\ ,
\end{equation}
where $C_n\in\mathbb R^M$ is the vector of listed option prices, $C\in\mathbb R^{M\times{\mathcal H}}$ is the matrix of modeled option prices, and $\phi_n\in\mathbb R^{\mathcal H}$ is a vector of probabilities. The inverse problem can be described succinctly as
\begin{equation}
\label{eq:disInverseProblem}
\min_{\phi\in\mathcal P_{\mathcal H}}\left\|C_n-C\phi\right\|^2\ ,
\end{equation}
where $\mathcal P_{\mathcal H}\doteq\{\phi\in\mathbb R^{\mathcal H}|\sum_j\phi^j=1,~\phi^j\geq 0~\forall j\}$ and $\|\cdot\|$ denotes the Euclidean norm. 

\begin{figure}[htbp] %  figure placement: here, top, bottom, or page
   \centering
   \includegraphics[width=6.5in]{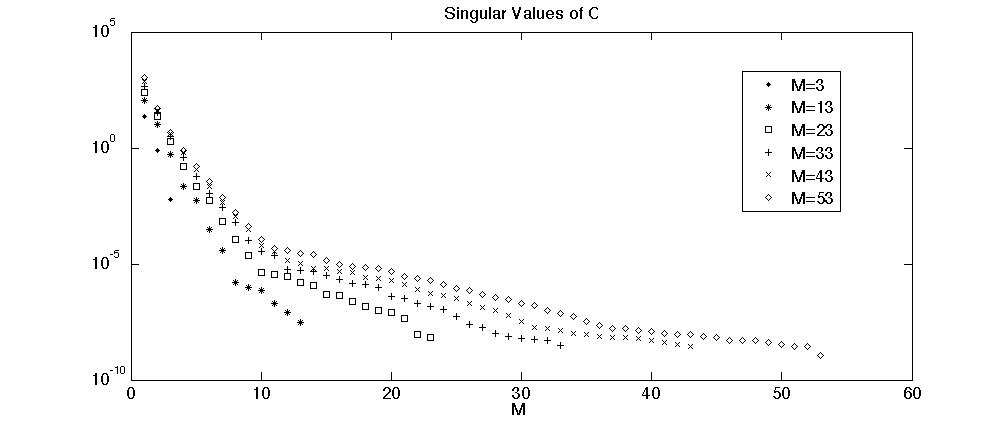} 
   \caption{\small For $N=M$ and $C\in\mathbb R^{M\times{\mathcal H}}$ generated by the Heston model, the singular values decay extremely quickly. The 1st `kink' in the rate of decay indicates that the effective dimension of $C$ is roughly 10 for all sizes. This also suggests that roughly 10 principle components capture much of the variation in options prices. When dealing with options data, this means that the added benefit of taking $M>10$ is relatively insignificant; the model fits `best' when the data consists of not much more than 10 options.}
   \label{fig:singularValues}
\end{figure}

Lagrange multipliers and linear programming can be used to solve equation \eqref{eq:disInverseProblem}, but the system should first be regularized because the matrix $C$ is usually ill-conditioned. For instance, given a fixed number of option prices, $C$ will quickly become ill-conditioned as the number of $x_i$'s is increased (i.e. $\mathcal H\gg M$ and the matrix is wider than it is tall). In fact, for $\mathcal H=M$ and $C$ composed of Heston model call prices, we see from Figure \ref{fig:singularValues} that $C$ still is ill-conditioned as the dimension grows. The `kinks' or `elbows' seen in Figure \ref{fig:singularValues} also suggest that the \textit{effective rank} of $C$ is roughly around 10 for all sizes, which means that roughly 10 principal components capture the majority of the variation. In other words, the added benefit of analyzing options data beyond $M=10$ is relatively insignificant.

Intuitively, one should be able to recognize the ill-conditioned nature of the problem by realizing that there is a marginal increase in the information gained by considering further refinements of the model matrix. For instance, all call-option prices are `hockey sticks', and after a certain point there is little to be learned by looking at more and more hockey sticks.

%%%%%%%%%%%%%%%%%%%%%%%%
\subsubsection{Tykhonov Regularization}
\label{sec:tykhonov}
Tykhonov regularizations are a tool that is often used to solve ill-posed systems. This regularization technique is similar to ridge regression (see \cite{hoerlKennard1970}). The idea is to add terms that penalize spiky behavior in the least-squares solution.  In a problem related to stochastic volatility, \cite{achdouPironneauBook} use Tykhonov regularizations to solve the inverse problem of estimating a local-volatility function. 

For some $\alpha>0$ and some degree of smoothness $d\geq 0$, a Tykhonov regularization involves solving a penalized least-squares problem,
\begin{equation}
\label{eq:LAreg}
\min_{\phi\in\mathcal P_{\mathcal H}}\left\{\|C_n - C\phi\|^2+\alpha\sum_{i=0}^d\|D^i\phi\|^2\right\}\ .
\end{equation}
where $(D\phi)_i=\frac{\phi_{i+1}-\phi_i}{\Delta x}$ (i.e. the finite difference between the indexed components of the vector). For example, we can take $d=0$ and then compute the singular value decomposition (SVD) of $C$ 
\[U\Sigma P^* = C\ ,\]
where $U\in\mathbb R^{M\times M}$ is an orthonormal matrix, $\Sigma\in\mathbb R^{M\times{\mathcal H}}$ is a banded matrix with the singular values of $A$ on its diagonal (and zeros everywhere else), and $P\in\mathbb R^{\mathcal H\times{\mathcal H}}$ is another orthonormal matrix. After computing the SVD, the inverse problem with $d=0$ can be written as a well-posed problem,
\begin{equation}
\label{eq:LAtykhonovReg}
\min_{\phi\in\mathcal P_{\mathcal H}}\|(\Sigma^*\Sigma+\alpha I)^{1/2}P^*\phi-(\Sigma^*\Sigma+\alpha I)^{-1/2}\Sigma^*U^*C_n\|^2\ ,
\end{equation}
with $I$ representing the $\mathcal H\times{\mathcal H}$ identity matrix. The linear system in equation \eqref{eq:LAtykhonovReg} is now $\mathcal H\times{\mathcal H}$, and the regularization parameter $\alpha$ insures that the problem is well posed. Judging from Figure \ref{fig:singularValues}, taking $\alpha = 10^{-4}$ for the Heston model produces a solution that considers 10 principle components of $C$, and the remaining singular values are increased to $\alpha$.

In general, regularized problems with $d=0,1,2\dots$ can be reduced to a least squares problem with linear constraints, and convexity arguments can be used to show that such problems have unique solutions. These solutions are consistent in the sense that they converge as $\mathcal H$ grows, but convergence of the solution to the true $\phi_n$ as $\mathcal H$ \textbf{and} $M$ grow is unlikely; there will always be some bias introduced by the Tykhonov regularization (as this is the `bias' component of the bias-variance trade-off associated with regularization).

Often times in statistics, the Tykhonov regularization is associated with the user's prior belief that some regularity is associated with the solution. Thus, the technique can be considered Bayesian. In this paper, our prior beliefs include the existence of a filtering density (see Proposition \ref{prop:SVMfilter}), and so a certain amount of entropy and smoothness can be imposed on the inverse problem. We write the regularized problem and label each piece as follows:
\begin{equation}
\label{eq:smoothProblem}
\min_{\phi\in\mathcal P}\left\{\underbrace{\|C_n-C\phi\|^2}_{\hbox{residual energy}}+\underbrace{\alpha\|\phi\|^2}_{\hbox{entropy}}+\underbrace{\beta\|D\phi\|^2}_{\hbox{smoothness}}\right\}\ .
\end{equation}
The `residual energy' is simply the magnitude of the residual vector, or the object that we seek to minimize. The `entropy' term penalizes $\phi$ with low entropy because minimizing the Euclidean norm raises the lower bound on entropy; from Jensen's inequality we see that $-\log\|\phi\|^2\leq -\left<\log\phi,\phi\right>=\hbox{entropy}(\phi)$. The `smoothness' term penalizes non-smooth behavior in $\phi$, which can (as we will see in Section \ref{sec:robustReg}) help to preserve the density's shape if there are small errors in $C$ caused by minor errors in the model parameters. %Obviously, there are problems for which regularizing terms will prevent the emergence of the true solution. For instance, the recovery of a $\delta$-function or point-mass becomes difficult (or even impossible) with regularization.

%%%%%%%%%%%%%%%%%%%
\subsection{Example: Inversion Through the Black-Scholes Model}
\label{sec:blackScholesExample}
Suppose that volatility remains constant over time, $\sigma_n\equiv \sigma(X_0)=X_0$, and that the market price of a European call option is the conditional expectation of the Black-Scholes price,

\begin{equation}
\label{eq:CnBlackScholes}
 C_n^i= \int\left(S_n\mathcal N(b_1^i(x))-K_ie^{-r(T-t_n)}\mathcal N(b_2^i(x))\right)\mathbb Q(X_0\in dx|\mathcal F_n^{\Delta t})\ ,
 \end{equation}
where $K_i$ is the strike price of the $i$th option, $\mathcal N(\cdot )$ is the CDF of a standard normal random variable, and
\begin{eqnarray}
\nonumber
b_1^i(x)&=&\frac{\log(S_n/K_i)+(r+.5x^2)(T-t_n)}{x\sqrt{T-t_n}}\ ,\\
\nonumber
b_2^i(x)&=&b_1(x)-x\sqrt{T-t_n}\ .
\end{eqnarray}
In this example we have simply
\[\phi_n(x)dx =  \mathbb Q(X_0\in dx|\mathcal F_n^{\Delta t})\ ,\]
from which we generate option prices using this risk-neutral expectation of the Black-Scholes price, and then compare the regularized solution to the true distribution. In the Black-Scholes theory it is assumed that options are priced with the true volatility parameter, which is equivalent to $\phi_n$ being a point-mass around the correct value, i.e. $\phi_n(x)= \delta_{\{\hat\sigma_{BS}=x\}}$, and which results in Black-Scholes implied volatility that has no smile. This example shows how a density $\phi_n$ with non-zero variance can produce an implied volatility smile.
\begin{figure}[htbp] %  figure placement: here, top, bottom, or page
   \centering
   \includegraphics[width=6.3in]{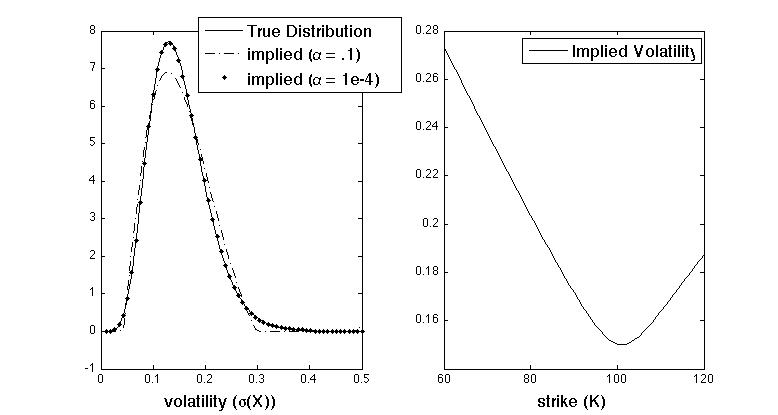} 
   \caption{\small \textbf{The Black-Scholes Example.} \textbf{Left:} In the Black-Scholes market with volatility being gamma distributed, the implied $\phi$ is very close when $\alpha=10^{-4}$. \textbf{Right:} The implied volatility, which exhibits a volatility skew that is similar to the skew that is observed from real market data. The additional convexity introduced by $\phi_n$ is good because it allows us to use the over-simplified Black-Scholes model, yet still have the smile that is associated with richer models.}
   \label{fig:BStoyExample}
\end{figure}

Given $\mathcal F_n^{\Delta t}$, suppose that $\sigma_n$ is gamma distributed, 
\[\phi_n(x) = \frac{x^{\nu-1}e^{-x/\zeta}}{\zeta^\nu\Gamma(\nu)}\qquad\forall n\geq 0\ ,\]
where $\nu=7.5$ and $\zeta=.02$. Then $\mathbb E^Q[\sigma(X_0)|\mathcal F_n^{\Delta t}] =\nu\zeta= .15$ and $\sqrt{var^Q(\sigma(X_0)|\mathcal F_n^{\Delta t})}=\sqrt{\nu\zeta^2}=.0548$. Suppose that $S_n=100$ and that we observed 61 call options with strikes $K_i =   59+i$ for $i=1,\dots,61$ with time to maturity $T-t_n=10/252$. We compute the Black-Scholes prices at the points $x_j = j\Delta x$ for $j=1,\dots,61$ where $\Delta x =.0082$, and then place them in the matrix $C\in\mathbb R^{61\times 61}$ given by

\[C^{ij}=S_n\mathcal N(b_1^i(x_j))-K_ie^{-r(T-t_n)}\mathcal N(b_2^i(x_j))\ ,\]
for all $i,j$,  and the regularized solution is
\[\widehat\phi_n = \arg\min_{\phi\in\mathcal P_{61}}\left\{\| C_n-C\phi\|^2+\alpha\|\phi\|^2\right\}\ ,\]
where $\mathcal P_{61} = \{\phi\geq 0:\sum_j\phi^j=1\}$. The matrix $C$ is neither full rank or well conditioned, as $rank(C)=28<61$ and $cond(C)\sim 10^{19}$.  However, $\widehat\phi_n$ is a good fit when the system is regularized with $\alpha =10^{-4}$, as
\[\| C_n-C\widehat\phi_n\|^2 \sim 10^{-10}\ ,\]
and $\left|.15-\sum_jx_j\widehat\phi_n(x_i)\right| \sim 10^{-6}$. The regularized solution is shown in the left-hand plot in Figure \ref{fig:BStoyExample}.

The implied volatility is shown in the right-hand plot of Figure \ref{fig:BStoyExample}. The at-the-money mark is $S_ne^{r(T-t_n)} = 100.9662$, which is at the low point of the smile and very close to the true average of volatility, $\hat\sigma_{BS}(K_{atm},t_n)\approx \mathbb E^Q[\sigma(X_0)|\mathcal F_n^{\Delta t}]=\zeta\nu  = .15$. The implied volatility also shows a smile that is similar to those of historical market data, but whose convexity and skew has been caused by $\phi_n$. This added convexity in the volatility smile is very similar to the theory proven by \cite{renaultTouzi1996}, and could be a useful tool when analyzing parsimonious models such as the Black-Scholes.

%%%%%%%%%%%%%%%%%%%
\subsection{Example: Inversion Through the Heston Model}
\label{sec:hestonToyExample}
\begin{figure}[htbp] %  figure placement: here, top, bottom, or page
   \centering
   \includegraphics[width=6.3in]{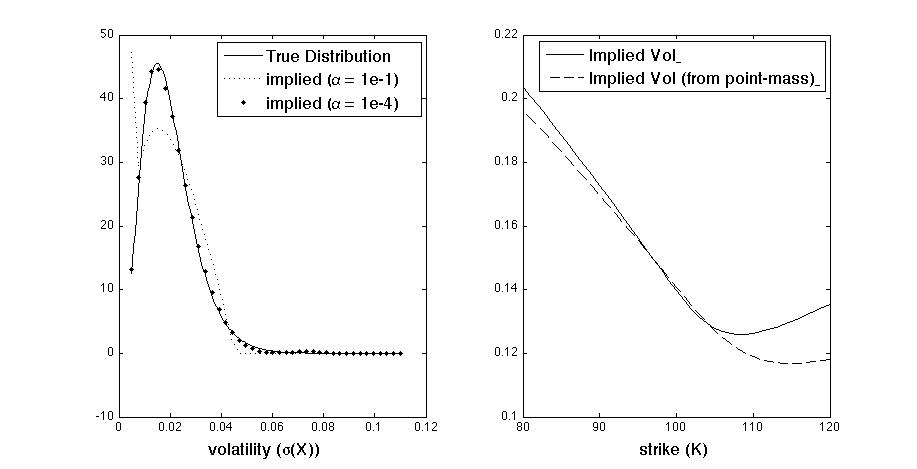} 
   \caption{\small \textbf{The Heston Example.} \textbf{Left:} The exact and implied $\phi_n$'s for a market where call-option prices are the conditional mean of their Heston model prices. The Tychonov regularization produces an accurate estimate of $\phi_n$ when $\alpha=10^{-4}$. \textbf{Right:} The implied volatility of $C_n=C\phi_n$ and the implied volatility of prices generated with a point-mass, $C\mathbf \delta_{\{\sigma_t^2=\int x\phi_n(dx)\}}$. We'll see in Section \ref{sec:hestonData} that the smile resulting from $\phi_n$ with non-zero variance improves the fit is similar to the smile of the SPX options.}
   \label{fig:hestonToyExample}
\end{figure}

Consider an example with mostly the same parameters as the Black-Scholes example in Section \ref{sec:blackScholesExample}, but with a Heston model,
\[dX_t = \kappa(\bar X-X_t)dt+\gamma\sqrt{X_t}dB_t^Q\ , \]
with volatility function $\sigma(X_t) = \sqrt{X_t}$, with parameter values $(\kappa,\bar X,\gamma,\rho)$=(2, .0225, .3, -.6), and by equation \eqref{eq:hestonVolRiskNeutral} we can (without loss of generality) take the volatility risk premium to be $\lambda=0$. At time $t_n$ we assume $X_n$ to be gamma distributed with density %(also its invariant distribution)
\[\mathbb Q(X_n\in dx|\mathcal F_n^{\Delta t}) =\phi_n(x)dx= \frac{x^{\nu-1}e^{-x/\zeta}}{\zeta^\nu\Gamma(\nu)}dx\ ,\]
where $\nu = .02/.005$ and $\zeta = .005$, so that $\mathbb E^Q[\sigma^2(X_n)|\mathcal F_n^{\Delta t}]=\nu\zeta=.02$ and $\sqrt{var^Q(\sigma^2(X_n)|\mathcal F_n^{\Delta t})} = \sqrt{\nu\zeta^2} =.01$, and we observe $41$ call options with strike prices $K_i = 79+i$ for $i=1,\dots,41$. Then, at $41$ points $x_j = j\Delta x$ for $j=1,2,\dots,41$ with $\Delta x=.0026$, we compute the Heston price of the call option with volatility $\sqrt{x_j}$ and strike $K_i$. The matrix $C\in\mathbb R^{41\times 41}$ is
\begin{equation}
\label{eq:star1}
C^{ij} = e^{-r(T-t_n)}\mathbb E^{Q}[(S_T-K_i)^+|S_n,X_n=x_j]\ , 
\end{equation}
with the right-hand side of equation \eqref{eq:star1} being a quadrature computation of the explicit call-option price that was originally derived in \cite{heston1993} (for numerical methods in computing the Heston formula, see \cite{carrMadan1999} and \cite{albrecherMayerSchoutensTistaert}). The left-hand plot in Figure \ref{fig:hestonToyExample} displays $\widehat\phi_n$ for this example, using a Tykhonov regularization of equation \eqref{eq:LAreg} with parameter $\alpha=10^{-4}$. The design matrix $C\in\mathbb R^{41\times41}$ has full rank as $rank(C)=41$, but has very high condition number (i.e. it is ill-conditioned) with $cond(C)=10^{11}$. After solving with regularization, the residual error is $\|C_n - C\widehat\phi_n\|^2\sim 10^{-7}$, and the solution's error is $\left| \sum_jx_j\widehat\phi_n(x_j) -.02\right|\sim 10^{-4}$.

The at-the-money implied volatility is approximately $.1367\approx.1414= \sqrt{.02}$, but is not as close to the true expected value as it was under the Black-Scholes model. Figure \ref{fig:hestonToyExample} shows the implied volatility generated by $C\phi_n$ along with the implied volatility generated by a point-mass distribution, $C\delta_{\{\sigma_t^2=\int x\phi_n(dx)\}}$. The figure illustrates how the Heston model generates a volatility smile: a non-zero value of the parameter $\rho$ produces smiles with skew, and an increase in the parameter $\gamma$ produces smiles that have more pronounced convexity. We see in Figure \ref{fig:hestonToyExample} that $\phi_n$ with non-zero variance makes the smile more pronounced. Indeed, we will see in Section \ref{sec:hestonData} that $\phi_n$ is helpful when fitting the Heston model to historical data.

This Heston model example is referred to in the analysis to come. So far, we have shown that the solution to the regularized problem obtained using the Heston model is an accurate estimate of the true distribution, but in the Section \ref{sec:error} we will see how accuracy of the solution is affected by parameter error and data imprecision.

%%%%%%%%%%%%%%%%%%%%%%%%%%%%%%%%%%%%%%%%%%%%%%%%%%%%%%%%%%%%%%%%%%%%%%%%%%%%%%%%%%%%%%%%%%%%%%%%%%%%%%
%%%%%%%%%%%%%%%%%%%%%%%%%%%%%%%%%%%%%%%%%%%%%%%%%%%%%%%%%%%%%%%%%%%%%%%%%%%%%%%%%%%%%%%%%%%%%%%%%%%%%%
\section{Error Analysis}\label{sec:error} Section \ref{sec:illPosedness} mentioned three sources of error encountered when solving the inverse problem. In this section we address two of them: parameter estimation error and measurement imprecision. The third type of error, which arises from numerical integration, we choose not to discuss because consistency of the approximated integrals is clear as $\mathcal H$ grows. All the analysis in this section is based on a model matrix $C$ that was built in the example of Section \ref{sec:hestonToyExample}; matrix $C$ whose entries are call prices under the Heston model.

%%%%%%%%%%%%%%%%%%%%%%%%%%%%%%%%%%%
\subsection{Measurement Imprecision and Over Fitting}
\label{sec:overfitting}
There is significant error introduced by the market's rounding of prices to the lowest denomination of currency. Theoretically, risk-neutral prices are real numbers in $\mathbb R$, but US-based exchanges quote these prices only to 2 decimal places (i.e. pennies are the lowest denomination of US currency). Thus, reducing the fitting error beyond the precision of the market results in over fitting. However, this additional error is the inverse problem's saving grace because it is no longer beneficial to take $\alpha$ vey small.

\begin{figure}[htbp]
\begin{minipage}[b]{1\linewidth}
   \centering
   \includegraphics[width=6.4in]{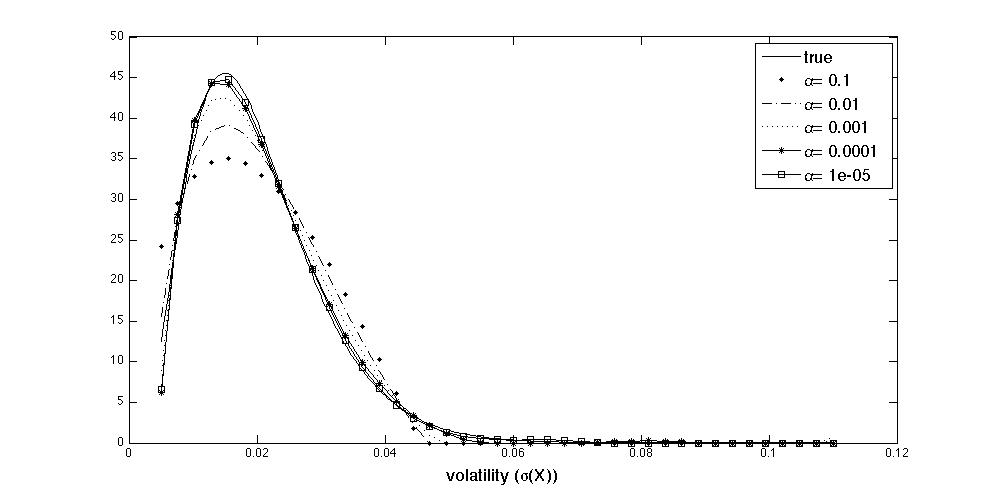} 
   \caption{\small Implied densities obtained using the Heston model, with measurements that have 16 decimal places of accuracy. Since the measurements are precise, the solution's error decreases as $\alpha$ decreases. In such a situation it is optimal to take $\alpha$ very small. }
   \label{fig:marginalimprove}

   \includegraphics[width=6.4in]{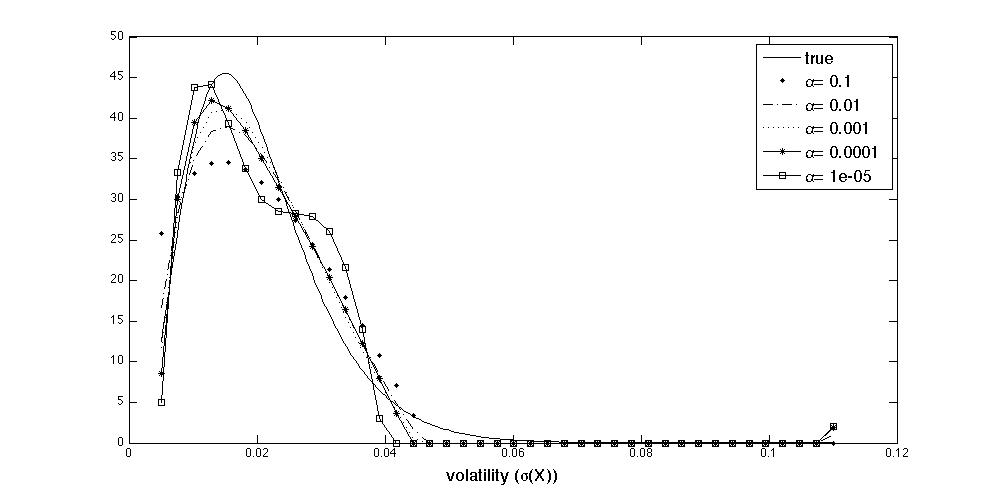} 
   \caption{\small Implied densiites obtained using the Heston model, with measurements that have 2 decimal places of accuracy. Since there is measurement imprecision, $\widehat\phi_n$'s shape starts to deteriorate at $\alpha =10^{-5}$ and thus the optimal $\alpha$ is greater than $10^{-5}$. By choosing $\alpha<10^{-5}$, we may be decreasing the residual, but the solution is over fitting the data rather than improving its error relative to the true solution.}
   \label{fig:marginalDecay}
   \end{minipage}
\end{figure}

To understand why smaller $\alpha$ is no longer beneficial in the presence of round-off error, consider the $L^\infty$-norm:

\[\|C_n-C\phi\|_\infty=\max_i\left|C_n^i-(C\phi)^i\right| \ .\]
If $C_n$ is truncated to 2 decimal places, it is over fitting if  the $L^\infty$-norm is minimized beyond an accuracy of order $10^{-3}$. In terms of the Euclidean norm, the potential for over fitting can be seen by using the inequality $\frac{1}{\sqrt {\mathcal H}}\leq\|\phi\|$ to obtain

\begin{equation}
\label{eq:star2}
\|C_n-C\phi\|_\infty^2+\frac{\alpha}{\mathcal H}\leq\|C_n-C\phi\|_\infty^2+\alpha\|\phi\|^2\leq\|C_n-C\phi\|^2+\alpha\|\phi\|^2\ .
\end{equation}
Thus, by taking very small $\alpha$ in equation \eqref{eq:star2} we see that the minimizing the Euclidean norm will result in reducing the $L^\infty$-norm, and hence there is over fitting. If prices are rounded to two decimal places, then $\alpha$ should be chosen as the maximum parameter for which the $L^\infty$-norm of the solution is less than $.005$
\[\alpha_0 = \max\{\alpha:\|C_n-C\widehat\phi_n\|_\infty< .005\}\ ,\]
(if the $L^\infty$-norm is less than $.005$ then all modeled prices are rounded to the recorded market prices, and so any further refinement of the residual is over fitting). In Figure \ref{fig:marginalimprove}, the experiment of Section \ref{sec:hestonToyExample} is repeated and the solution to the inverse problem is shown for decreasing $\alpha$'s when there is very little measurement error. The measurements in $C_n$ are accurate to 16 decimal places, and there is improvement in the fit as we decrease $\alpha$. However, in Figure \ref{fig:marginalDecay} we show the solutions for decreasing $\alpha$'s when measurements in $C_n$ are accurate to only 2 decimals, and it appears that the solution begins to deteriorate for values of $\alpha$ that are less than $10^{-3}$. From Table \ref{tab:measurementErrors} we see that it is in fact the case that $\alpha=10^{-3}$ is the approximate cut-off before over fitting begins.

\begin{table}[htbp]
\caption{\small \textbf{$\mathbf{L^\infty}$ Error of Fit for Various $\alpha$'s and Measurement Precision}}
\begin{center}
\begin{tabular}{l|c|c}
\label{tab:measurementErrors}
&\multicolumn{2}{c}{measurement precision}\\
$\alpha$&16 decimals&2 decimals\\
\hline
$10^{-1}$&$.0047$&$.009$\\
$10^{-2}$&$.0016$&$.0057$\\
$10^{-3}$&$10^{-4}$&$.005$\\
$10^{-4}$&$10^{-5}$&$.005$\\
$10^{-5}$&$10^{-6}$&$.005$
\end{tabular}
\end{center}
\end{table}

\subsection{Robustness to Parameter Uncertainty}
\label{sec:robustness}
Section \ref{sec:inverseProblem} has explained a model-indifferent method for computing $\phi_n$ on the latent state of volatility. However, we have not yet addressed the issue(s) of parameter estimation and the associated estimation error. The reality is that parameters are sometimes more important in derivative pricing than the process itself. For instance, long-term options are more sensitive to the long-time average of volatility than they are to the initial value. Thus, a good deal of emphasis on the sensitivity to parameter uncertainty is warranted.

For a vector-valued parameter $\theta$, the modeled option price is denoted by $C^i(t_n,S_n,x;\theta)$, and we solve the inverse problem
\[\min_{\phi\in\mathcal P_{\mathcal H}}\left\{\|C_n-C(\theta)\phi\|^2 +\alpha\|\phi\|^2\right\}\ ,\]
where matrix $C(\theta)$ has entries $C^{ij}(\theta) = C^i(t_n,S_n,x_j;\theta)$. For this particular parameter value, the solution to this system is 
\[\widehat\phi_n(\theta)\doteq\arg\min_{\phi\in\mathcal P_{\mathcal H}}\left\{\|C_n-C(\theta)\phi\|^2 +\alpha\|\phi\|^2\right\}\ ,\]
and we are interested in how corrupt $\widehat\phi_n(\theta)$ becomes as the parameter value strays from the true parameter. For instance, the parameter of the Heston model from Section \ref{sec:hestonToyExample} is $\theta=(\kappa,\bar X,\gamma,\rho)$, and we want to estimate the solution's error as these 4 factors change.

\begin{figure}[htbp] %  figure placement: here, top, bottom, or page
   \centering
   \includegraphics[width=6.4in]{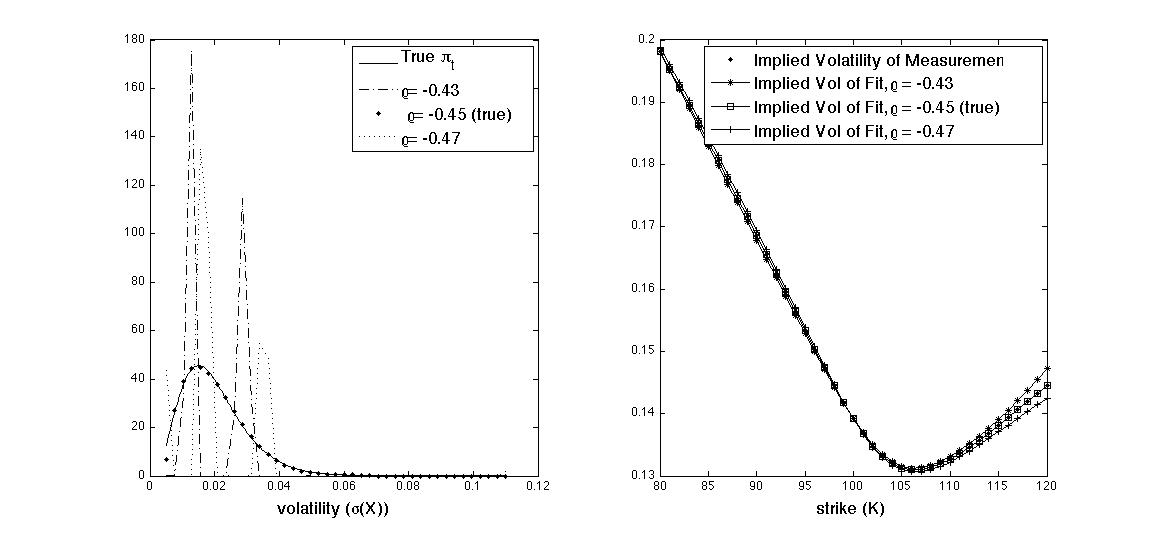} 
   \caption{\small \textbf{With Precise Measurements, the Solution Error Caused by Parameter Uncertainty.} \textbf{Left:} For the Heston model with $\alpha=10^{-6}$, a slight change in the parameter $\rho$ prevents an accurate recovery of the density's shape. \textbf{Right:} The implied volatility of $C_n$ along with the implied volatility of $C(\theta)\widehat\phi_n(\theta)$ computed with various $\rho$. Notice the implied volatility for $\rho=.-43$ or $-.47$ has significant error on the right tail.}
   \label{fig:perturbed}
\end{figure}

To illustrate the sensitivity of the solution to perturbations in $\rho$, we repeat the experiment from Section \ref{sec:hestonToyExample} with all the same numbers except with $\alpha=10^{-6}$ and with a little bit of movement in $\rho$. Figure \ref{fig:perturbed} shows how the solution is very close to the true distribution when the correct parameter is used, but then is easily corrupted by perturbations to $\rho$.

The particular method used for parameter estimation is not the issue here, but rather the solution's error that is caused by parameter uncertainty. Section \ref{sec:robustReg} presents a study whose aim is to characterize the changes in $\widehat\phi_n(\theta)$ that occur due to changes in the parameterization of the inverse problem; in particular changes that are within a small neighborhood of the `true' value of $\theta$.
%%%%%%%%%%%%%%%%%%%%%%%%%%%%%%%
\subsubsection{Robustness Via Additional Regularization}
\label{sec:robustReg}
By adding a smoothness term, we are assuming the distribution on $\sigma_n$ is continuous \textit{and differentiable}. Solutions benefit from this assumption because spiky behavior like that seen in Figure \ref{fig:perturbed} are penalized. This section makes a study out of the example in Figure \ref{fig:perturbed}. The study shows how additional regularity can provide qualitative improvements, but higher order moments and the residual $\|C_n-C(\theta)\widehat\phi_n(\theta)\|$ indicate that robustness has not been achieved. However, we also see that when there is measurement imprecision, there is not much improvement in the residual when we use the true parameters. In particular, measurement imprecision and parameter uncertainty leads to estimates of higher order moments that cannot be relied upon, but 1st and 2nd moments are reliable and the shape of fit can also be considered `good.'

\begin{figure}[htbp] %  figure placement: here, top, bottom, or page
   \centering
   \includegraphics[width=6.4in]{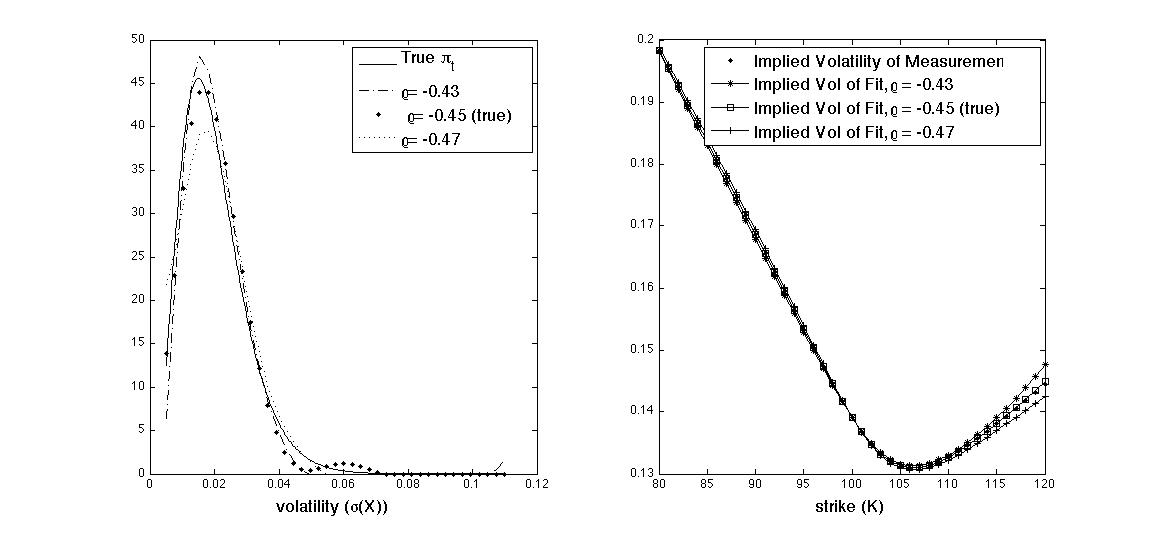} 
   \caption{\small \textbf{With Precise Measurements, the Addition of a Smoothness Term to Provide Robustness Against Parameter Uncertainty.} \textbf{Left:} Using a smoothness term with $\beta=10^{-8}$, we repeat the experiment that was done to produce Figure \ref{fig:perturbed}. The smoothness term preserves $\phi_n$'s shape, whereas the case $\beta=0$ shown in Figure \ref{fig:perturbed} exhibits spiky behavior from only a small change in $\rho$. \textbf{Right:} However, the solution's error is still seen in the implied volatility --even with the smoothness term.}
   \label{fig:robust}
\end{figure}

Consider the inverse problem,
\[\min_{\phi\in\mathcal P_{\mathcal H}}\left\{\|C_n-C(\theta)\phi\|^2+\alpha\|\phi\|^2+\beta\|D\phi\|^2\right\}\ ,\] 
with $\alpha=10^{-6}$ and $\beta = 10^{-8}$, with $C$ constructed from the Heston model, and with $C_n$ accurate to within 16 decimal places. Figure \ref{fig:robust} shows the results that are shown in Figure \ref{fig:perturbed}, but with the smoothness term added. There appears to be some qualitative improvement, but the implied volatilities in Figure \ref{fig:robust} look very much like those in Figure \ref{fig:perturbed}, indicating that perhaps the smoothness has made only cosmetic improvements in the implied density.

Let's now shift our attention to some quantitative results in the study, namely the implied density's moments and the residuals. Recall from Section \ref{sec:hestonToyExample} that the true distribution was a gamma, with parameters $\nu=.02/.005 $ and $\zeta=.005$; the true moments are those listed in Table \ref{tab:posteriorMoments}. For each parameterization we've computed an estimate of $\phi_n$, from which we compute the implied density's moments, and then compare these moments to their respective counterpart in Table \ref{tab:posteriorMoments}. These results are shown in Table \ref{tab:moments}, along with the residual. It is clear from Table \ref{tab:moments} that 1st and 2nd moments are robust to parameter uncertainty, but that higher order moments suffer, both when there is smoothing and without. The table also shows us that the residual is considerably lower when the correct parameter is used and $\beta=0$, which indicates that parameter uncertainty can have a big effect, and that smoothness can introduce more bias and not contribute toward greater accuracy. Similar studies can be done for the other parameters of the Heston model (see Table \ref{tab:momentGamma} for the Heston model's sensitivity to perturbations in $\gamma$; there is also a lack of robustness). 
\begin{table}[!h]
\caption{\small \textbf{Volatility's Moments when $\phi_n$ is a gamma density with parameters $\nu=.02/.005$ and $\zeta=.005$}}
\begin{center}
\begin{tabular}{c|c}
\label{tab:posteriorMoments}
moment&value\\
\hline
mean&.02\\
standard dev&.01\\
skew&1\\
kurtosis&4.5
\end{tabular}
\end{center}
\end{table}

\begin{table}[!h]
\caption{\small \textbf{Implied Filter's Moments and Residuals, with Perturbations in $\mathbf\rho$ (Heston Simulation)}}
\begin{center}
\begin{tabular}{c|c|c|c|c|c|c|c}
\label{tab:moments}
&\multicolumn{3}{c|}{no smoothing ($\mathbf\alpha=10^{-6}$, $\beta=0$)}&\multicolumn{3}{c|}{with smoothing ($\mathbf\alpha=10^{-6}$, $\beta=10^{-8}$)}\\
&$\rho=-.43$&$\rho=-.45$ (true)&$\rho=-.47$&$\rho=-.43$&$\rho=-.45$ (true)&$\rho=-.47$\\
\hline
mean&0.0203&0.0203&0.0203&0.0203&0.0203&0.0203\\
std dev&0.0105& 0.0099&0.0098&0.0105&0.0099&0.0098\\
skew&3.4097& 1.0582&0.4736&3.3994 &1.2709&0.5515\\
kurtosis&29.4659&4.6610&2.1620&28.2152& 6.1667&2.8774\\
residual&$10^{-4}$&$10^{-11}$&$10^{-4}$&$10^{-4}$&$10^{-7}$&$10^{-4}$
\end{tabular}
\end{center}
\end{table}

\begin{table}[!h]
\caption{\small \textbf{Implied Filter's Moments and Residuals, with Perturbations in $\mathbf\gamma$ (Heston Simulation)}}
\begin{center}
\begin{tabular}{c|c|c|c|c|c|c|c}
\label{tab:momentGamma}
&\multicolumn{3}{c|}{no smoothing ($\mathbf\alpha=10^{-6}$, $\beta=0$)}&\multicolumn{3}{c|}{with smoothing ($\mathbf\alpha=10^{-6}$, $\beta=10^{-8}$)}\\
&$\gamma = .28$&$\gamma=.3$ (true)&$\gamma = .32$&$\gamma = .28$&$\gamma=.3$ (true)&$\gamma = .32$\\
\hline
mean&0.0202&0.0203&0.0203&0.0202&0.0203&0.0203\\
std dev&0.0115&0.0099&0.0089&0.0115& 0.0099&0.0089\\
skew&3.5664&1.0582&0.1748&3.4699&1.2709&0.2167\\
kurtosis&28.5255&4.6610&1.8650&26.7772&6.1667&2.3007\\
residual&$10^{-4}$&$10^{-11}$&$10^{-4}$&$10^{-4}$&$10^{-7}$&$10^{-4}$
\end{tabular}
\end{center}
\end{table}

\begin{figure}[htbp] %  figure placement: here, top, bottom, or page
   \centering
   \includegraphics[width=6.4in]{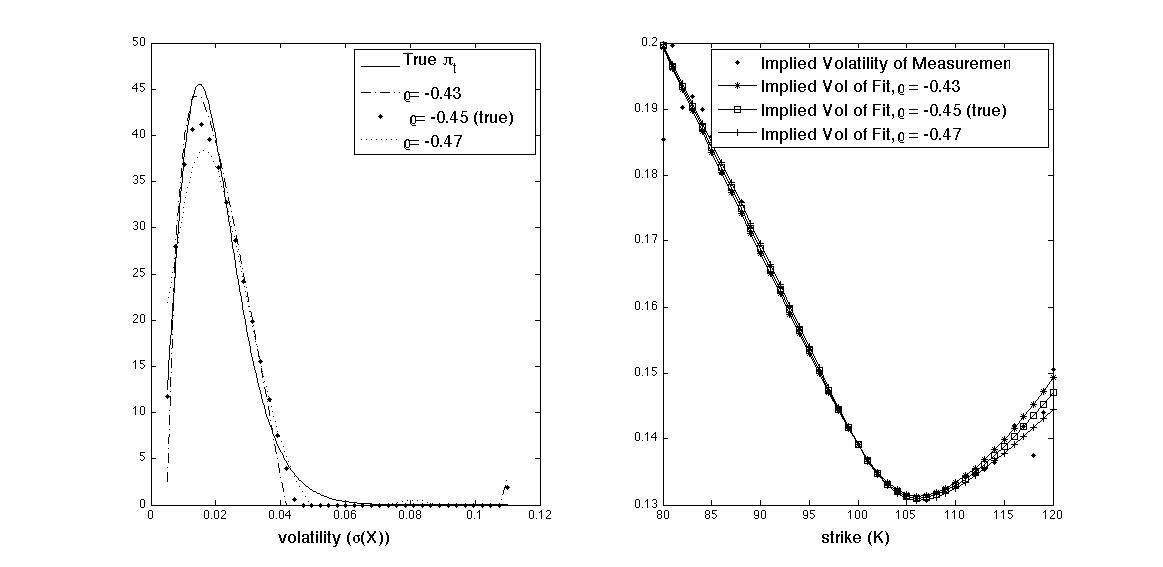} 
   \caption{\small \textbf{With Measurement Imprecision, the Solution Error with Parameter Uncertainty.} \textbf{Left:} Implied densities from the Heston model when there is both parameter uncertainty and measurement imprecision. We know from Section \ref{sec:illPosedness} to take $\alpha=10^{-3}$ with the Heston model, and in doing so we have mitigated the error that would have occurred due to parameter uncertainty had $\alpha$ been smaller. \textbf{Right:} The right tail of the data's implied volatility is now scattered around the smile curves of all 3 parameter values, indicating that measurement imprecision has made the problem somewhat insensitive to parameter uncertainty.}
   \label{fig:perturbMeasErr}
\end{figure}

Now, let's add measurement imprecision to the study. It was shown in Section \ref{sec:overfitting} that one should use $\alpha=10^{-3}$ when measurement precision is 2 decimals, and from Figure \ref{fig:perturbMeasErr} we see some preservation of the distribution's shape when $\alpha=10^{-3}$ and $\beta=0$, but from Table \ref{tab:perturbMeasErr} we see that 3rd and 4th moments of volatility are still erroneous. However, the table also shows us far less disparity in the residual under parameter uncertainty, indicating that the fit is not as sensitive to parameter uncertainty as it was with accurate measurements.
\begin{table}[!h]
\caption{\small \textbf{Implied Filter's Moments and Residuals, with Perturbations in $\mathbf\rho$ and Measurement Imprecision (Heston Simulation $\mathbf\alpha=10^{-3}$ $\mathbf\beta=0$)}}
\begin{center}
\begin{tabular}{c|c|c|c}
\label{tab:perturbMeasErr}
&$\rho=-.43$&$\rho=-.45$ (true)&$\rho=-.47$\\
\hline
mean&0.0203&0.0203&0.0204\\

std dev& 0.0113&0.0109&0.0104\\
skew&3.8259& 2.8589&1.2094\\
kurtosis&30.6764&23.3747&7.2838\\
residual&$10^{-4}$&$10^{-6}$&$10^{-5}$
\end{tabular}
\end{center}
\end{table}

To summarize: the evidence from Figures \ref{fig:perturbed}, \ref{fig:robust}, and Tables \ref{tab:moments} and \ref{tab:momentGamma} indicate that the inverse problem is not robust to parameter uncertainty. We clearly see that (i) perturbations in $\rho$ affects the recovery of $\phi_n$ under the Heston model, (ii) that 1st and 2nd moments remain mostly intact, (iii) that error occurs in the higher order moments, (iv) that smoothness in the regularization does little to correct error in higher moments, and (v) the residual is significantly lower when the parameter is correct. However, Figure \ref{fig:perturbMeasErr} and Table \ref{tab:perturbMeasErr} show that measurement-imprecision in the data mitigates these concerns because the improved residual that results from the true parameter is hardly as dramatic as it is when data is precise.

%%%%%%%%%%%%%%%%%%%%%%%%%%%%%%%%%%%%%%%%%%%%%%%%%%%%%%%%%%%%%%%%%%%%%%%%%%%%%%%%%%%%%%%%%%%%%%%%%%%%%%%
%%%%%%%%%%%%%%%%%%%%%%%%%%%%%%%%%%%%%%%%%%%%%%%%%%%%%%%%%%%%%%%%%%%%%%%%%%%%%%%%%%%%%%%%%%%%%%%%%%%%%%%
\section{Application to SPX Data}
\label{sec:data}
We now solve the inverse problem given call and put options on the S\&P 500 (SPX). We use daily SPX data from 2005, consisting of the index's closing price and the bid-ask spreads on European call and put options with expiries of 1 month, 2 months, 3 months, 6 months, 1 year, 1.5 years, 2 years and 3 years. The options expire on the Saturday after the 3rd Friday of each month, and we choose to discard options with less than 7 business days to maturity. Thus, there is a `reset' in the data that occurs roughly 1 week after the start of each month, which causes periodic breaks or `maturity cycles' in the parameter estimates and sometimes in $\widehat\phi_n$. 

Often times, the problem of fitting a stochastic-volatility model is divided into three parts: $K$, $T$, and $t$, where
\begin{itemize}
\item `$K$' is the problem of fitting a single cross-section of the implied-volatility smile across different strikes,
\item `$T$' or `big-t',  is the problem of fitting across multiple maturities,
\item `$t$' or `little-t', is the problem of modeling changes in the implied-volatility surface over time.
\end{itemize}
In this section, we fit the model and solve the inverse problem across both $K$ and $T$, separately for each day, using both a Heston model and a Heston model with jumps to construct the matrix $C$. When we use the Heston model, there are little-t effects that appear clearly in the implied $\widehat\phi_n$, there are little-t effects in the estimated model parameters, and the estimated 30-day variance-swap rate is consistent with the VIX time series. The little-t effects in $\widehat\phi_n$ appear in the form of periodic behavior that is consistent with the monthly maturities of the options. In particular, we suspect that these effects are a periodic premium that get implied into the Radon-Nykodim derivative $\Lambda_n=\frac{\phi_n}{\pi_n}$ introduced in Definition \ref{def:separable}; it is more than likely that $\Lambda_n$ contains the periodic component of $\widehat\phi_n$ because the physical filtering density $\pi_n$ is computed from the time series of stock prices that are not monthly periodic (recall Remark \ref{rem:riskPrem}). When we include jumps, the maturity cycles fade and are only noticeable in the estimated jump intensity. Furthermore, the jump models has an estimated variance-swap rate that is less consistent with the VIX time series. In general, we conclude that the parsimonious parameterization of the Heston model leads to fits that are easier to interpret, and that the model is less prone to over fitting. We find it harder to draw conclusions from the Heston model with jumps because there are so many parameters that it is difficult to interpret the implied objects, and it is possible that over fitting has occurred.

The following is a summary of the contents in the coming section: we solve the inverse problem using the Heston model and also solve using a Heston model with jumps. We find that the density $\phi_n$ provides extra explanatory power, as its inclusion results in a better fit to the S\&P500 options data than the simpler fit that assumes $X_n$ is observed. We use the implied density $\widehat\phi_n$ as a proxy for $\mathbb Q(~\cdot~ |\mathcal F_n^{\Delta t})$ and observe maturity effects in the mean and standard deviations from the density $\widehat\phi_n$ that was implied by the Heston model. When we add jumps, $\widehat\phi_n$ does not exhibit maturity effects, and the only little-t effects we can spot are those that appear vaguely in the estimated jump-intensity parameter. Comparatively, we find the Heston model without jumps to be in better-sync with the VIX data over a longer period, whereas the jump model exhibits some erratic behavior that could be a result of over fitting. We conclude that the adage of `\textit{sparser is better}' applies here, as we find fits of the no-jump Heston model are easier to interpret, and we find it difficult to interpret the fitted parameters and implied densities from the Heston model with jumps.
%%%%%%%%%%%%%%%%%%%%%%%

\subsection{Results Using the Heston Model \& Heston Model with Jumps}
\label{sec:hestonData}

We consider the same Heston model with jumps that was considered by \cite{BCC1997},
\begin{equation}
\label{eq:hestonModel}
\begin{array}{ccl}
dS_t/S_{t^-}&=&rdt+\sqrt{X_t}\left(\rho dB_t^Q+\sqrt{1-\rho^2}dW_t^Q\right)+J_td\mathcal Z_t-\nu dt\ ,\\
dX_t&=&\kappa(\bar X-X_t)dt+\gamma\sqrt{X_t}dB_t^Q\ ,
\end{array}
\end{equation}
where $B_t^Q$ and $W_t^Q$ are independent Brownian motions under the $\mathbb Q$-pricing measure, $1+J_t$ is a log-normal random variable with parameters $\mu_J>-1$ and $\sigma_J>0$ such that
\begin{align*}
\mathbb E^Q\log(1+J_t)&=\log(1+\mu_J)-.5\sigma_J^2\ ,\\
var^Q(1+J_t) &= \sigma_J^2\ .
\end{align*}
The process $\mathcal Z_t$ is an independent Poisson process with intensity $\lambda_J\geq 0$ and with $d\mathcal Z_t=\mathcal Z_t-\mathcal Z_{t^-}$, and $\nu$ is a compensator. The compensator is included to insure that discounted returns are a martingale, which requires $\nu$ to be chosen so that the following equality holds:
\[r=\frac{1}{t}\mathbb E^Q\int_0^t\frac{dS_\tau}{S_\tau} = r-\nu+\frac{1}{t}\mathbb E^Q\left\{\int_0^t\sqrt{X_\tau}\left(\rho dB_\tau^Q+\sqrt{1-\rho^2}dW_\tau^Q\right)+\int_0^tJ_\tau d\mathcal Z_\tau\right\}  \]

\[=r-\nu+\frac{\lambda_J}{t}\int_0^t\mathbb E^QJ_\tau d\tau =r-\nu+\frac{\lambda_J}{t}\int_0^t(\mathbb E^Qe^{\log(1+J_\tau)} -1)d\tau =r-\nu+\lambda_J\mu_J\ ,\]
which implies $\nu = \mu_J\lambda_J$ (see \cite{BCC1997, carrWu2009,lewis} for further details on these types of models with jumps). We also enforce the Feller condition in the diffusion process, $\gamma^2\leq 2\kappa\bar X$. The explicit pricing formula for call options under this model is given in \cite{BCC1997}.

For each day we take a weighted sum of the bid and ask prices on call and put options, and then estimate the model parameters by finding the minimizer to the following residual: 
\begin{equation}
\label{eq:star3}
\min_{x_0,\theta}\sum_i\omega_i\left|\frac{C_n^{i,ask}+C_n^{i,bid}}{2}-C^i(t_n,S_n,x_0;\theta)\right|^2\ ,
\end{equation}
where $C^i(t_n,S_n,x_0;\theta) = e^{-r(T_i-t)}\mathbb E^Q[(S_{T_i}-K_i)^+|S_n,\sqrt{X_n}=x_0]$\footnote{The parameter $r$ is part of a term structure of discount rates, and includes an adjustment for the SPX's dividend rate. For each maturity time $T$, the discount rate $r_{t,T}$ is computed from the put-call parity, on which we run a linear regression for parameters $a_1$ and $a_2$ over options of maturity $T$. The optimal parameters fit the data to the strike prices, $K$, to the model $P_t^{T,ask}(K)-C_n^{T,bid}(K)+S_t=a_1K+a_2\mathbf 1_{\{ask-bid\}}$, where $C_n^{T,\cdot}$ and $P_t^{T,\cdot}$ are the market's price of calls and puts (respectively) with maturity $T$, and $\mathbf 1_{\{ask-bid\}}$ is the indicator that we have taken the difference $P_t^{T,ask}-C_n^{T,bid}$ as opposed to $P_t^{T,bid}-C_n^{T,ask}$. The discount rate is $r_{t,T} = -\log(\hat a_1)/(T-t)$ where $\hat a_1$ is the least squares fit; this estimator usually has very low variance if $T-t$ is greater than a week.} with model parameter $\theta$, and $\omega_i$ is a weight-factor that assigns importance according to moneyness and the bid-ask spread, 
\[\omega_i \propto \frac{\exp\{-(10\log(e^{-r(T-t)}K_i/S_n))^2\}}{\hat\sigma_{BS}(C_n^{i,ask})-\hat\sigma_{BS}(C_n^{i,bid})}\ .\] 
The minimization problem in equation \eqref{eq:star3} is similar to the inverse problem of Section \ref{sec:inverseProblemPosed}, but has effectively assumed that $\phi_n$ is a point-mass at $x_0$ (minimization of equation \eqref{eq:star3} is the procedure referred to in the stochastic volatility literature as calibration). Under the model that we've defined, the distribution of $S_T$ has an affine characteristic function, so we compute $C_n^i(t_n,S_n,x_0;\theta)$ with the inverse Fourier transform, both in the case with jumps and without (numerical methods for computing options prices with Fourier transforms are described in \cite{carrMadan1999}, \cite{albrecherMayerSchoutensTistaert}, and with jumps in \cite{lewis}). The parameter estimates that we obtain from the 2005 SPX data are shown in Tables \ref{tab:hestonParams} and \ref{tab:hestonJumpParams}, and they are within reasonable distance to the estimates in \cite{BCC1997} and \cite{eraker2004} (they used data from different dates so their estimates should not be exactly the same).

\begin{figure}[htbp]
\begin{minipage}[b]{0.48\linewidth}
\centering
\includegraphics[scale=.6]{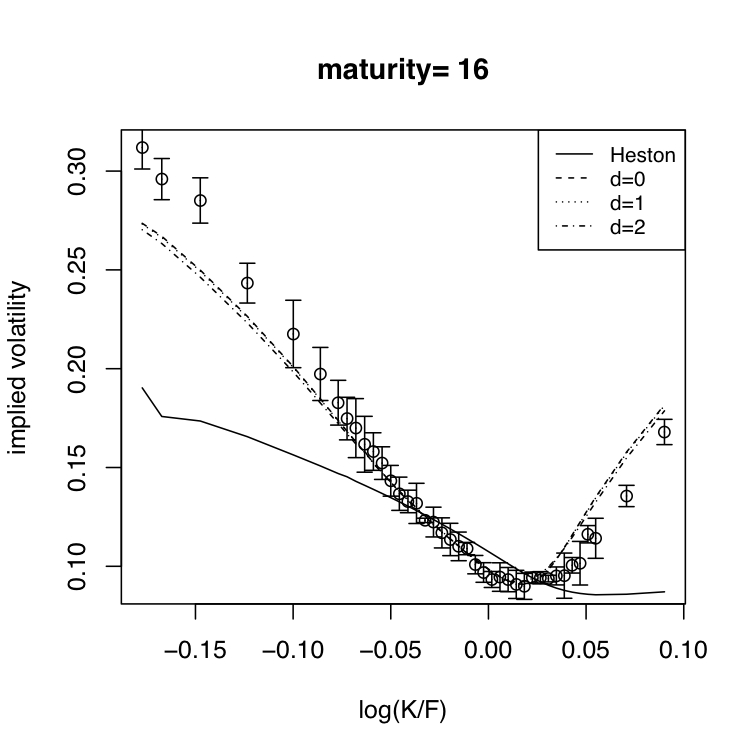}

\includegraphics[scale=.6]{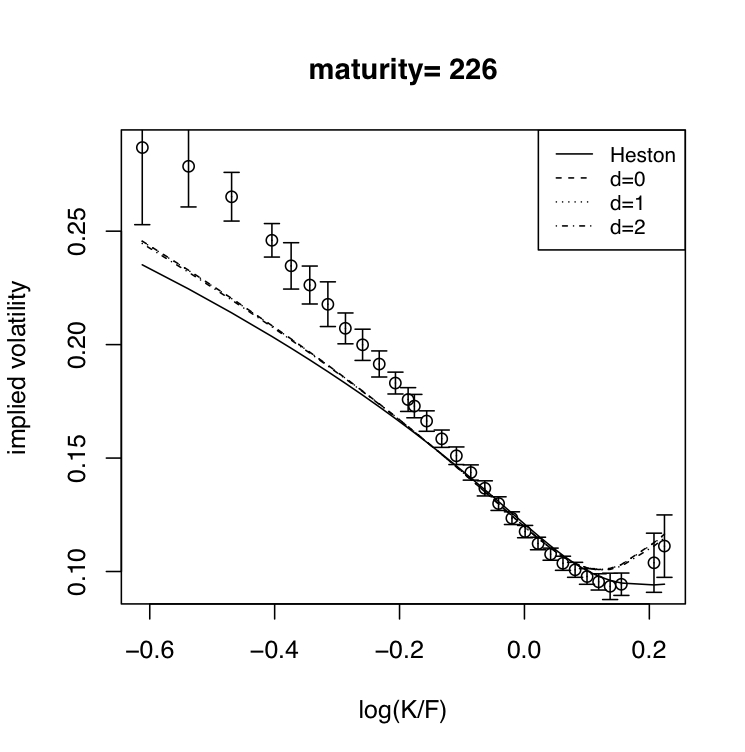}

\end{minipage}
\hspace{0.5cm}
\begin{minipage}[b]{0.45\linewidth}
\centering

\includegraphics[scale=.6]{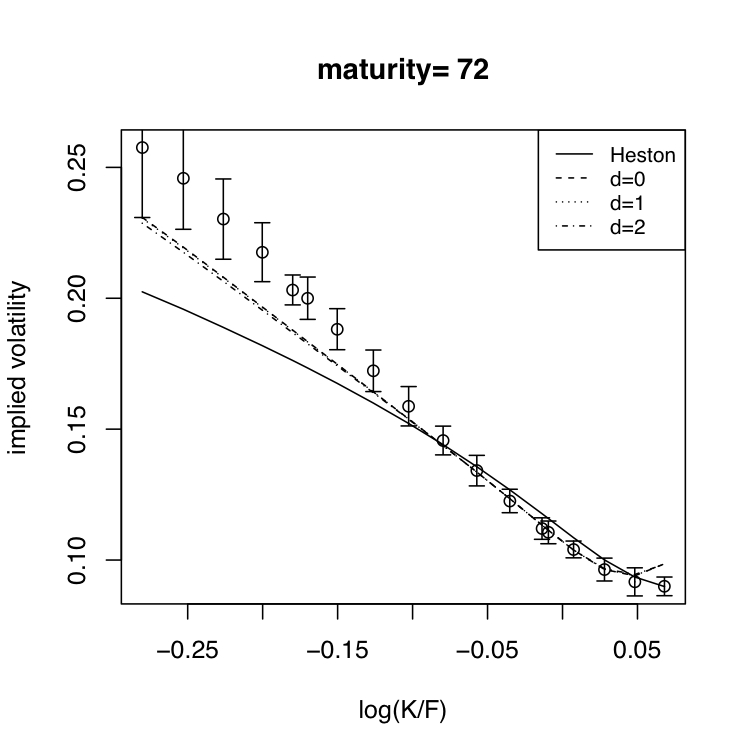}

\includegraphics[scale=.6]{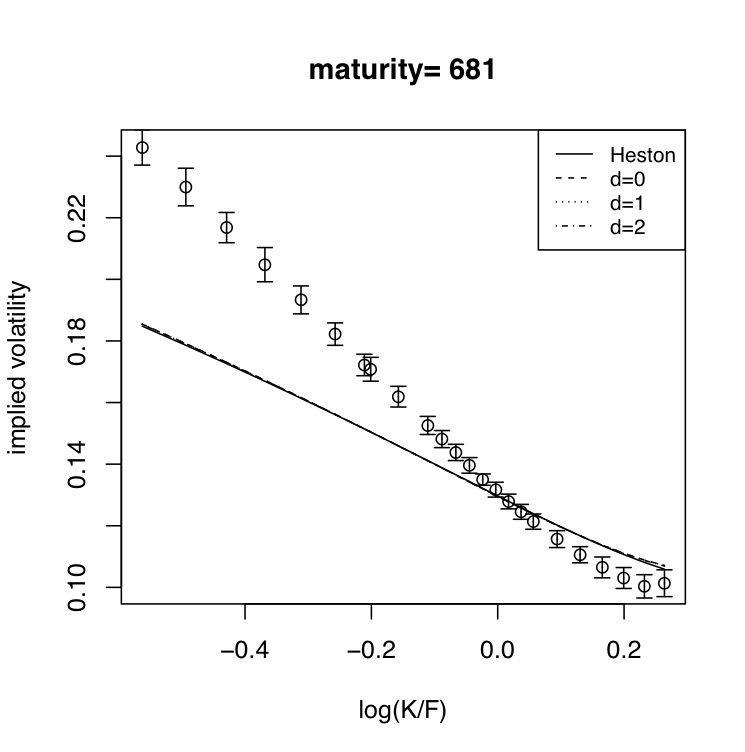}

\end{minipage}
\caption{\small The implied volatility of the Heston model fit on February 3, 2005. `Heston' refers to $\hat\sigma_{BS}(C(x_0))$, and `d=0, 1, 2' refer to $\hat\sigma_{BS}(C\widehat\phi_n)$ with $\widehat\phi_n$ computed with $d$ degrees of regularization. The error bar is the bid-ask spread, with the top of the brace being the ask price and the bottom being the bid. An important thing to notice in these plots is that $\phi_n$ has helped the Heston model to fit the short time-to-maturity options, which is an indication that volatility uncertainty may be of concern when the market is pricing options with short time-to-maturity.}
\label{fig:ivol}
\end{figure}

\begin{figure}[htbp]
\begin{minipage}[b]{0.48\linewidth}
\centering
\includegraphics[scale=.6]{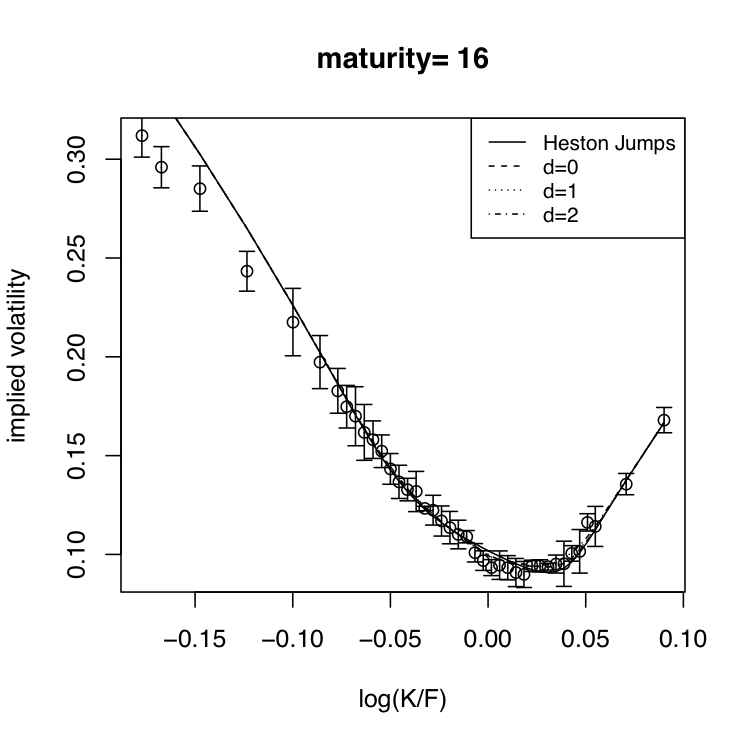}

\includegraphics[scale=.6]{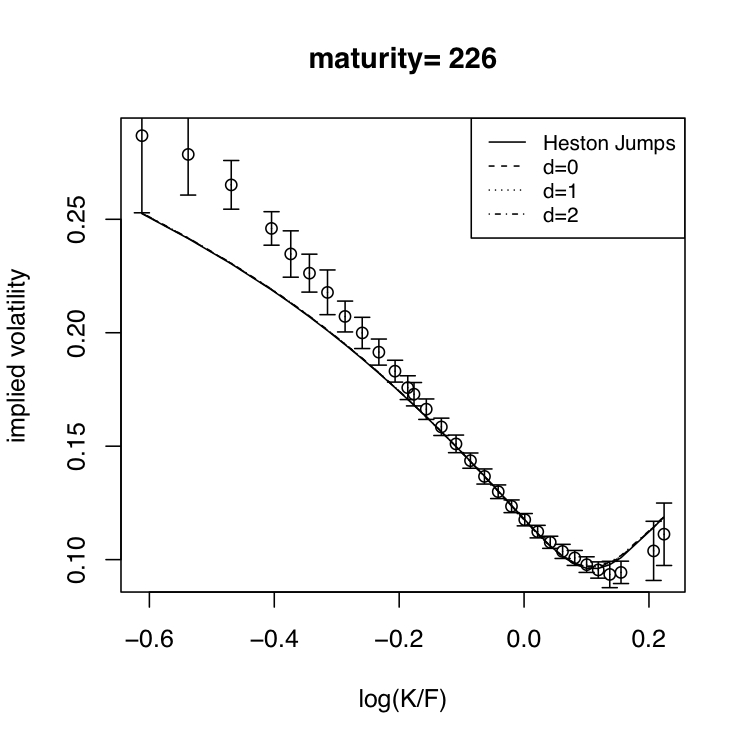}

\end{minipage}
\hspace{0.5cm}
\begin{minipage}[b]{0.45\linewidth}
\centering
\includegraphics[scale=.6]{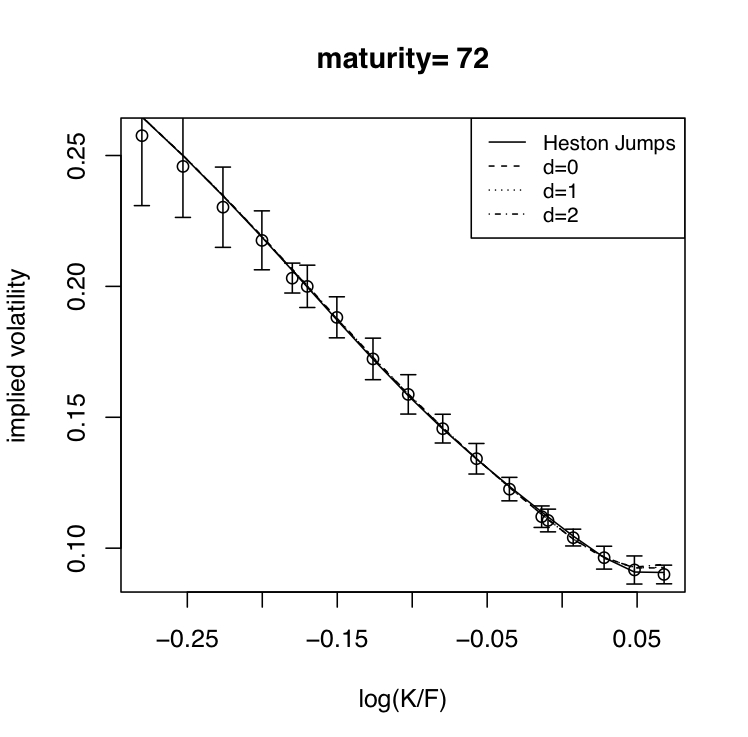}

\includegraphics[scale=.6]{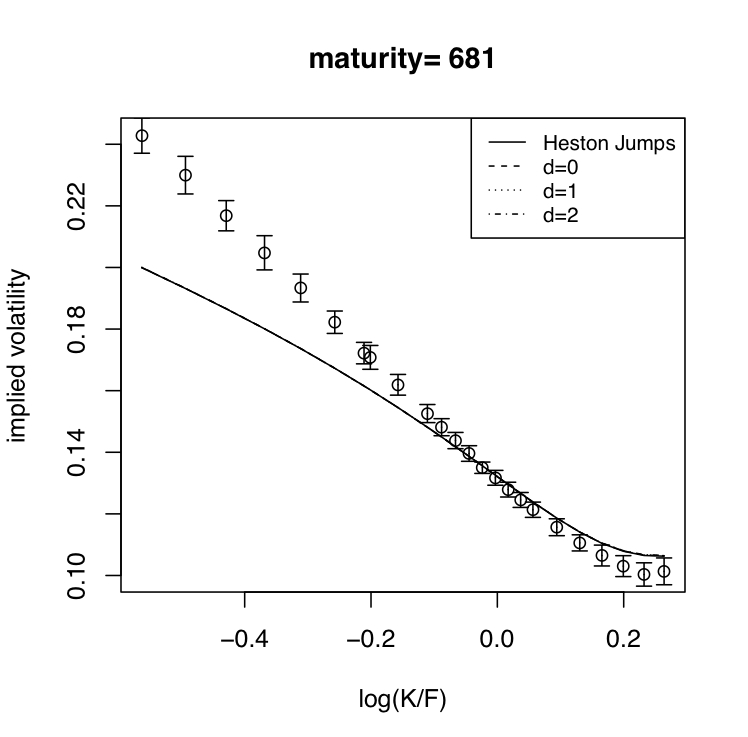}

\end{minipage}
\caption{\small The implied volatility of the Heston model fit with jumps on February 3, 2005. `Heston Jumps' refers to $\hat\sigma_{BS}(C(x_0))$, and `d=0, 1, 2' refer to $\hat\sigma_{BS}(C\widehat\phi_n)$ with $\widehat\phi_n$ computed with $d$ degrees of regularization. The error bar is the bid-ask spread, with the top of the brace being the ask price and the bottom being the bid. By adding jumps we have made the Heston model significantly richer, and thus there are enough degrees of freedom to allow the model to fit option prices pretty well. Thus the $\widehat\phi_n$ obtained using the jump model has very low variance, and is close to a point-mass. }
\label{fig:ivolJump}
\end{figure}

For cases both with and without jumps, we solve the inverse problem 
 \[\min_{\phi\in\mathcal P_{\mathcal H}}\left\{\|C_n-C\phi\|_\omega^2+\alpha_0\|\phi\|^2+\alpha_1\|D\phi\|^2+\alpha_2\|D^2\phi\|^2\right\}\ ,\]
 where $\mathcal P_{\mathcal H}=\{\phi\in\mathbb R^{\mathcal H}:\phi^i\geq 0,~\sum_i\phi^i=1\}$, $\|\cdot\|_\omega$ denotes the weighted Euclidean norm with the same weights that were used to estimate parameters, and the $\alpha$'s are chosen based on the desired degree of regularization (see Table \ref{tab:alphas}).
 \begin{table}[!h]
\caption{\small \textbf{Values of $\mathbf\alpha$ for Different Degrees of Regularity}}
\begin{center}
\begin{tabular}{c|c|c|c}
\label{tab:alphas}
$d$&$\alpha_0$&$\alpha_1$&$\alpha_2$\\
\hline
0&$10^{-3}$&0&0\\
1&$10^{-3}$&$10^{-7}$&0\\
2&$10^{-3}$&$10^{-7}$&$10^{-11}$
\end{tabular}
\end{center}
\end{table}
For each day, our analysis is based on four probability measures: the point-mass that is centered at $x_0$ estimated by equation \eqref{eq:star3}, and the three solutions computed using Tykhonov regularizations with $d=0,1,2$. Figures \ref{fig:ivol} and \ref{fig:ivolJump} show the implied volatility of the bid-ask spread and the fitted option prices. Notice in Figure \ref{fig:ivol} how added convexity from $\phi_n$ improves the fit of the Heston model for shorter time-to-maturity, but the modeled prices do not fit so well for longer time-to-maturity. In Figure \ref{fig:ivolJump} we see an improved fit across all maturities when we include jumps, but long maturities still do not fit perfectly and $\phi_n$ appears to have little effect on the fit. It is mentioned in \cite{FPS2011} that stochastic-volatility models can have difficulty describing both short and long time-to-maturity with a single parameterization. However $\phi_n$'s improvements to the fits for Heston model without jumps indicates that perhaps volatility uncertainty is of concern when the market is pricing options with shorter time-to-maturity. 
  
\begin{figure}[htbp]
\begin{minipage}[b]{0.45\linewidth}
\centering
\includegraphics[scale=.6]{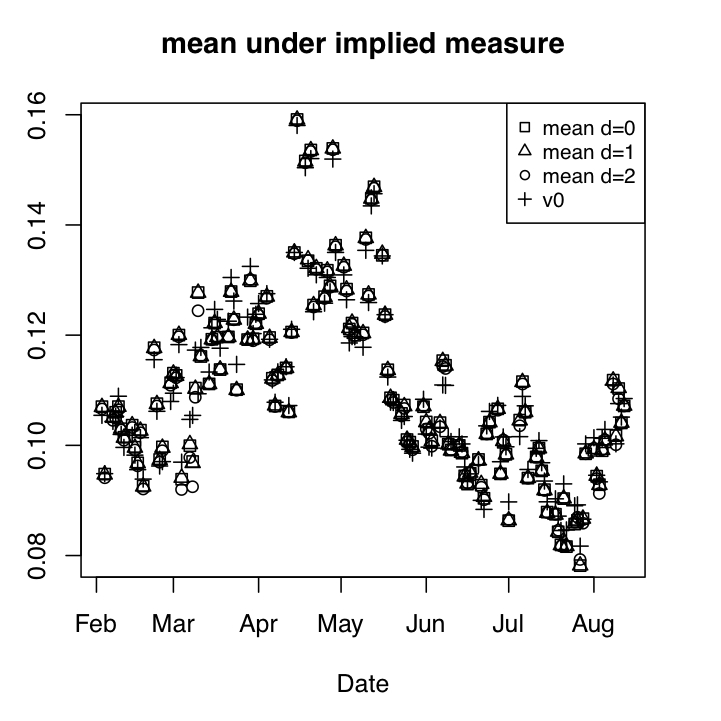}

\includegraphics[scale=.6]{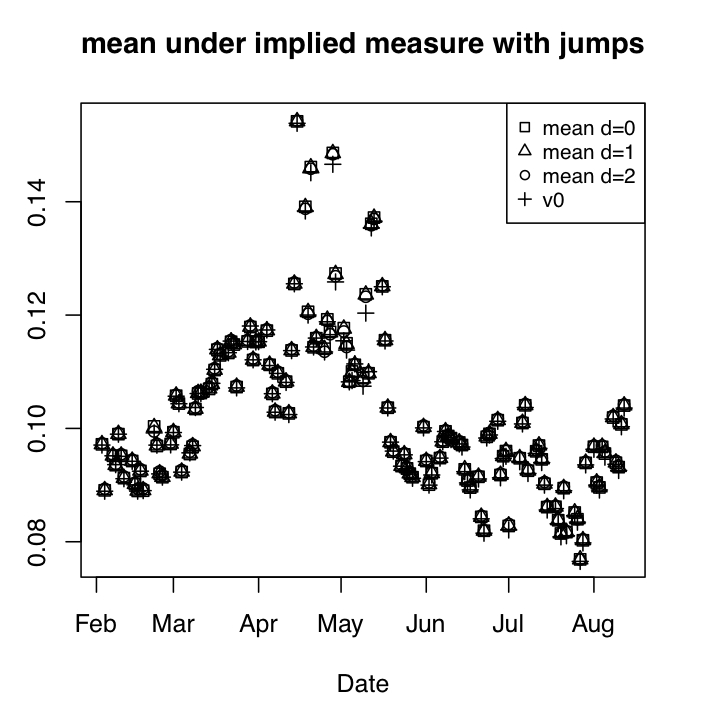}
%\caption{\small default}
%\label{fig:figure2}
\end{minipage}
\hspace{0.5cm}
\begin{minipage}[b]{0.45\linewidth}
\centering
\includegraphics[scale=.6]{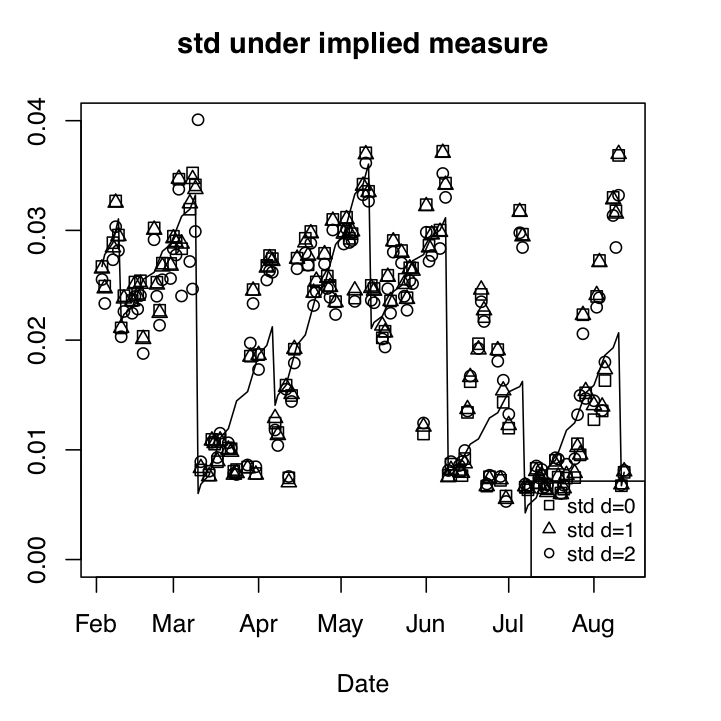}
%\caption{\small default}
%\label{fig:figure3}
\includegraphics[scale=.6]{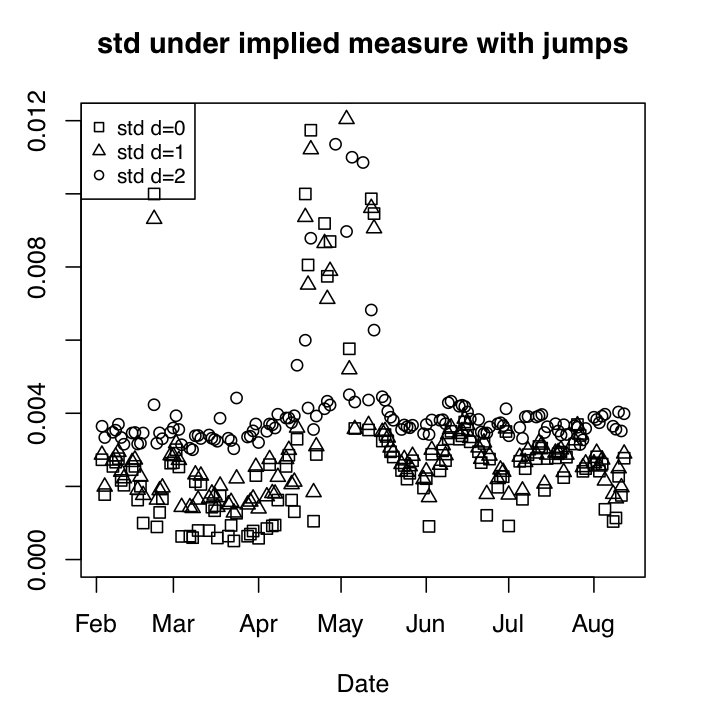}
%\caption{\small default}
%\label{fig:figure4}
\end{minipage}
\caption{\small Using $\widehat\phi_n$ as a proxy for $\mathbb Q(~\cdot~|\mathcal F_n^{\Delta t})$, the figure shows the time evolution of the risk-neutral filtering moments for the SPX data from 2/3/2005 to 8/12/2005. \textbf{Top left:} $\sqrt{\mathbb E^Q[X_n|\mathcal F_n^{\Delta t}]}$ as estimated from the implied density obtained from the Heston model without jumps. \textbf{Top Right:} $\sqrt{var^Q(X_n|\mathcal F_n^{\Delta t})}$ as estimated from the implied density obtained from the Heston model without jumps. Notice that we have drawn in the least absolute deviations (LAD) fit between maturity dates to highlight the increasing filter variance of $X_n$ as time to maturity decreases. \textbf{Bottom Left:} $\sqrt{\mathbb E^Q[X_n|\mathcal F_n^{\Delta t}]}$ as estimated from the implied density obtained from the Heston model \textbf{with} jumps. \textbf{Bottom right:} $\sqrt{var^Q(X_n|\mathcal F_n^{\Delta t})}$ as estimated from the implied density obtained from the Heston model \textbf{with} jumps.}
\label{fig:hestonMeanStd}
\end{figure}

We saw in Section \ref{sec:robustReg} how the implied density's 1st and 2nd moments were reasonably accurate, so we pay close attention to these moments and how they evolve from day to day; they are shown in Figure \ref{fig:hestonMeanStd} for the various $\widehat\phi_n$'s that we've computed. Notice how the implied density's standard deviation under the Heston model has some variation, but the implied density under the model with jumps has standard error that is very small. Also notice that we've identified the maturity cycles in the implied density's standard deviation when using the Heston model, which we highlight with a solid line in the upper-right plot of Figure \ref{fig:hestonMeanStd}. The solid line is the least absolute deviations (LAD) fit between maturity dates.

To summarize, $\widehat\phi_n$ appears consistent across for the two models and with varying degrees of regularity. The Heston without jumps produces a fit that benefits from the additional convexity effects of $\phi_n$. Also, the model without jumps exhibits maturity cycles, which is important because it means that the model is not implying spurious homogeneity. In contrast, the Heston model with jumps produces an improved fit, but produces $\widehat\phi_n$'s that have very little variance and no little-t effects. The apparent lack of little-t effects raises some skepticism, and as we see in the next section, it is hard to see the little-t effects in the jump model's estimated parameters as well. Comparatively, it appears that the sparser model (i.e. the Heston without jumps) produces implied filters that have simple and intuitive explanations.
%%%%%%%%%%%%%%%%%%%%%%%

\subsection{Maturity Cycles and `Little-t' Behavior}
\label{sec:matCycles}

It was observed in \cite{FPS2004} that a dynamic parameterization of a stochastic-volatility model was able to account for maturity cycles and stabilize the parameters. In our experiments, we've re-estimated the parameters everyday so as to give the stochastic-volatility model the best possible chance to capture all the little-t effects, yet we still see little-t effects in $\widehat\phi_n$ from the inverse problem. In particular, when we solve the inverse problem using the Heston model with no jumps, there is periodic and increasing standard deviation of $X_n$ under the distribution $\widehat\phi_n$ (see the upper-right plot in Figure \ref{fig:hestonMeanStd}). These periodic increases in standard deviation indicate that the risk premium associated with volatility uncertainty increases as time-to-maturity decreases. Since $\pi_n$ is computed using stock prices that have no maturity effects, if we assume separability as defined in Definition \ref{def:separable}, then we strongly suspect that $\Lambda_n \doteq \frac{\phi_n}{\pi_n}$ (as defined in  Proposition \ref{prop:iteratedExpectations}) is the periodic component and that it represents a maturity-dependent risk premium on volatility uncertainty in the options market. The presence of maturity cycles is a fact in the data, and so an absence of little-t effects in the fitted parameters and/or $\widehat\phi_n$ would mean that they have been absorbed by the multiple degrees of freedom provided by the model. Such an absorption of the little-t effects seems to be happening when we include jumps with the Heston model, and suggests the possibility of over fitting.

The maturity cycles in the upper-right plot of Figure \ref{fig:hestonMeanStd} occur around 1 week into the new month, which corresponds to 5 business days prior to option expiration, and so we are certain that these periodic effects are caused by the departure of shorter time-to-maturity data (i.e. options for which $T-t_n$ is close to zero). Indeed, options for which $T-t_n$ is small often imply higher volatility-of-volatility and have a more exaggerated smile, and so the parameter $\gamma$ is often estimated to be higher at times near to maturity. Thus, by discarding options for which $T-t_n$ is small, we are letting go of the data that implies higher volatility-of-volatility, and so there is a drop in volatility-of-volatility on the day that we discard this data.%\footnote{Apart from SPX options, there are some options markets where there is effectively a continuum of expiries, and the data does not exhibit these same maturity cycles.}

\begin{figure}[htbp]
\begin{minipage}[b]{0.48\linewidth}
\centering
\includegraphics[scale=.6]{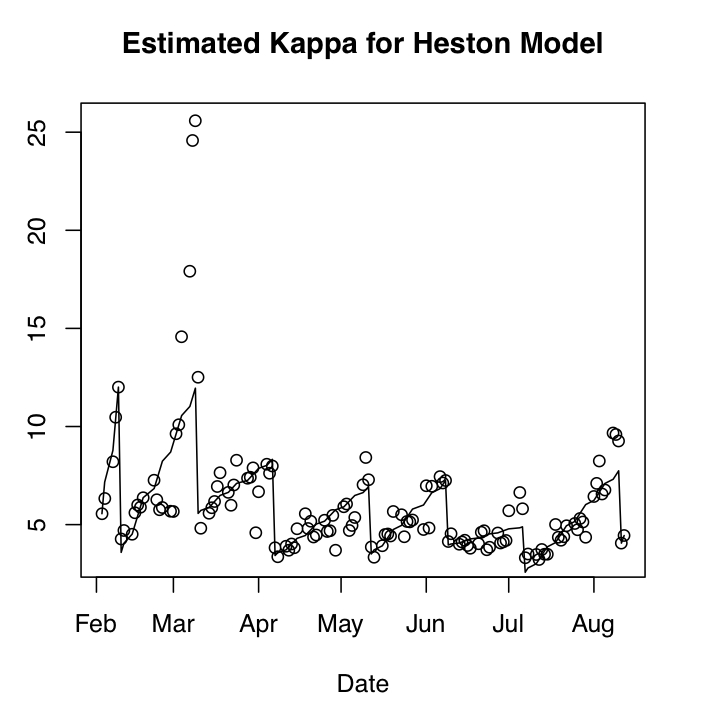}

\end{minipage}
\hspace{0.5cm}
\begin{minipage}[b]{0.45\linewidth}
\centering
\includegraphics[scale=.6]{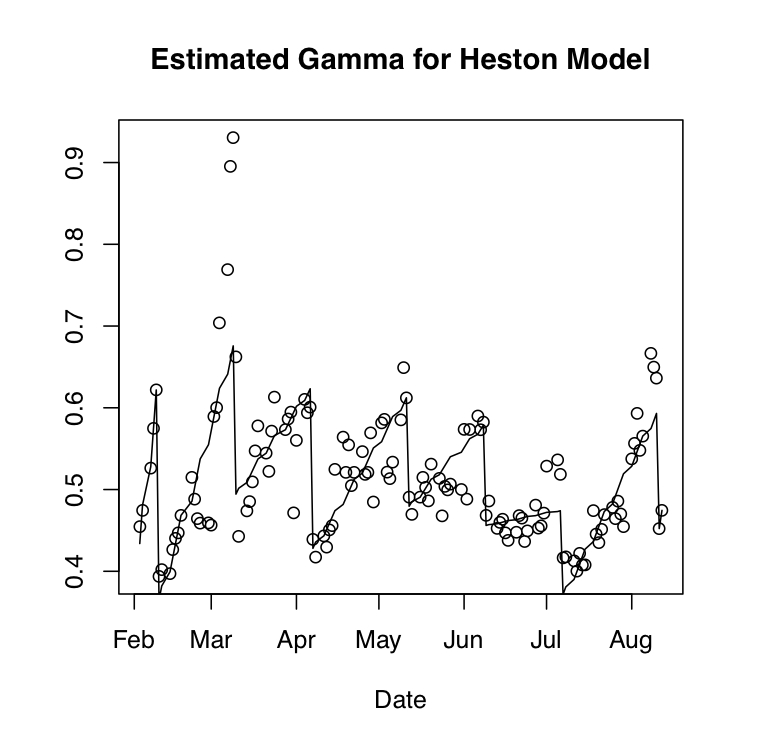}

\end{minipage}
\caption{\small \textbf{Left:} Maturity cycles in the daily estimates of $\kappa$. \textbf{Right:} Maturity cycles in the daily estimates of $\gamma$. In both plots, the solid line is the least absolute deviations (LAD) fit between maturities. The reason that $\kappa$ exhibits maturity cycles is because we have enforced the Feller condition, which means that an increase in volatility-of-volatility often requires an increase in $\kappa$.}
\label{fig:hestonParamsEst}
\end{figure}

\begin{figure}[htbp] %  figure placement: here, top, bottom, or page
   \centering
   \includegraphics[scale=.6]{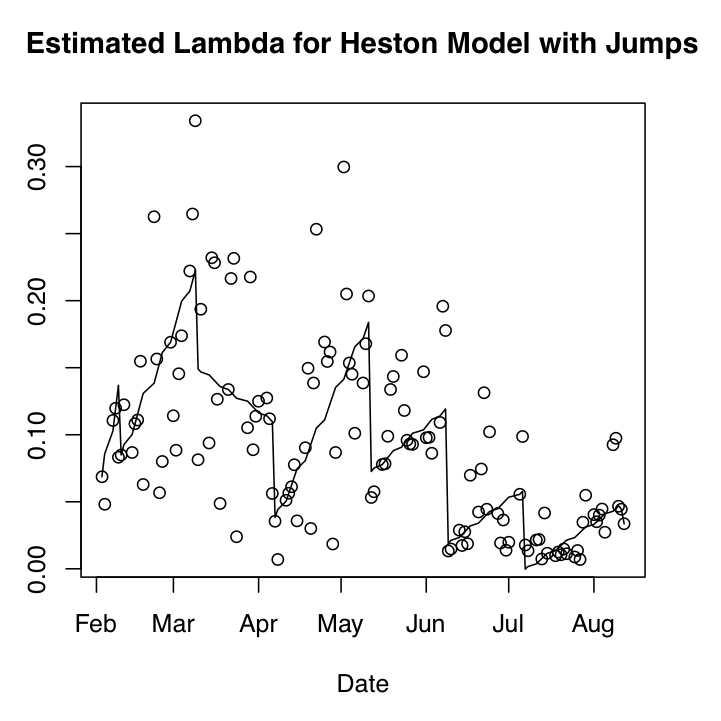} 
   \caption{\small The maturity cycles that occur in the estimated $\lambda_J$ in the Heston model with jumps. These maturity cycles are not as strong as those for the model without jumps, and they are the only maturity cycles that are visible within the jump model. This lack of visibility indicates that perhaps over fitting has occurred.}
   \label{fig:hestonJumpParamsEst}
\end{figure}

The maturity cycles in the fitted Heston model's volatility-of-volatility can be seen further in Figure \ref{fig:hestonParamsEst}, where the daily estimates of $\kappa$ and $\gamma$ are shown along with an LAD fit to highlight the rate at which the parameters increase. It should be remarked that $\kappa$ exhibits a maturity cycle in-part because we've enforced the Feller condition in parameter estimation; $\gamma^2\leq 2\bar X\kappa$ and so periodic increases in $\gamma$ result in periodic increases in $\kappa$. 

When we include jumps, the maturity cycles disappear in the both $\widehat\phi_n$ and the diffusion parameters. However, the fitted values of the jump intensity $\lambda_J$ exhibit some evidence of maturity cycles, as we see in Figure \ref{fig:hestonJumpParamsEst}, but these cycles are not as prevalent as those in Figures \ref{fig:hestonMeanStd} or \ref{fig:hestonParamsEst}. Intuitively, it makes sense that increases in the jump intensity would handle little-t effects caused by short time-to-maturity, because jumps have a lot of explanatory power for options with short time-to-maturity. Since we know that little-t behavior exists and is present in the data, it must be implied into the fit in some way or another. Therefore, we must assume that maturity cycles are present in the jump model's fitted parameters and/or $\widehat\phi_n$, but that we are simply not able to see them. Thus, the jump model has buried the maturity cycles, and perhaps over fitting has occurred.

%%%%%%%%%%%%%%%%%%%%%%%
 \subsection{Comparison With VIX}
Another important piece of analysis is the comparison of our fitted stochastic volatility to the VIX. For a fixed time period of length $\tau^*>0$, let $QV_{t,\tau^*}$ denote quadratic variation,
\[QV_{t,\tau^*} \doteq \lim_{|p|\searrow0}\sum_{t_i\in[t,t+\tau^*)}\log(S_{t_{i+1}}/S_{t_i})^2=\int_t^{t+\tau^*}X_sds+\sum_{i=\mathcal Z_t}^{\mathcal Z_{t+\tau^*}}\left(\log(1+J_i)\right)^2\ ,\]
where the limit holds in probability as $|p|\doteq\sup_i(t_{i+1}-t_i)\rightarrow 0$. Quadratic variation divided by $\tau^*$ is the continuous time analogue of the floating-leg for a variance swap contract, and from equation \eqref{eq:hestonModel} and using $\widehat\phi_n$ as a proxy for the risk-neutral filter, we have the swap rate:
 \begin{align}
\nonumber
 &\frac{1}{\tau^*}\mathbb E^Q[QV_{t_n,\tau^*}|\mathcal F_n^{\Delta t}]\\
  \label{eq:QV}
 &=\underbrace{\bar X+\frac{\mathbb E^Q[X_n|\mathcal F_n^{\Delta t}]-\bar X}{\kappa \tau^*}\left(1-e^{-\kappa \tau^*}\right)}_{\hbox{diffusion component}}+\underbrace{\lambda_J\left(\sigma_J^2+(\log(1+\mu_J)-.5\sigma_J^2)^2\right)}_{\hbox{jump component}}\\
 \label{eq:proxyQV}
&\approx\bar X+\frac{\int x\widehat\phi_n(x)dx-\bar X}{\kappa \tau^*}\left(1-e^{-\kappa \tau^*}\right)+\lambda_J\left(\sigma_J^2+(\log(1+\mu_J)-.5\sigma_J^2)^2\right)\ ,
\end{align}
which sets to zero the initial cost of entry into a variance swap. Obviously, the diffusion component in equation \eqref{eq:QV} is by itself the swap rate under the Heston model (without jumps), and \eqref{eq:QV} in it's entirety is the swap rate in the presence of jumps.

For variance swaps on SPX with $\tau^*=$ 30 days, a useful tool for computing the risk-neutral swap rate is the VIX index. The VIX is a volatility gauge that is computed from SPX options, and is related to the risk-neutral prediction of the 30-day variance. \cite{derman} derived the VIX formula for computing the fixed-leg of the variance swap when $dS_t$ has no jump component,

\[ VIX_t = 100\sqrt{\frac{2}{\tau^*}\left(\int_{F_{t,\tau^*}}^\infty\frac{\mathbb E^Q[(S_{t+\tau^*}-K)^+|\mathcal F_t^{\Delta t}]}{K^2}dK+\int_0^{F_{t,\tau^*}}\frac{\mathbb E^Q[(K-S_{t+\tau^*})^+|\mathcal F_t^{\Delta t}]}{K^2}dK\right)}\ ,\]
where $F_{t,\tau^*}$ is the forward rate on $S_{t+\tau^*}$ at time $t$, and the integrands are given by the market prices of European call and put options. It has since been shown in \cite{carrWu2009} that in the presence of jumps, the risk-neutral swap rate is proportional to VIX squared plus a jump term,
\begin{align}
\nonumber
&\frac{1}{\tau^*}\mathbb E^Q[QV_{t_n,\tau^*}|\mathcal F_n^{\Delta t}] \\
\nonumber
&= (.01\times VIX_n)^2 +\frac{2}{\tau^*}\mathbb E^Q\left[\sum_{i=\mathcal Z_n}^{\mathcal Z_{t_n+\tau^*}}.5(\log(1+J_i))^2+\log(1+J_i)-J_i\Bigg|\mathcal F_n^{\Delta t}\right]\\
\nonumber
&= (.01\times VIX_n)^2 +2\lambda_J\left(\frac 12\left(\sigma_J^2+(\log(1+\mu_J)-.5\sigma_J^2)^2\right)+\log(1+\mu_J)-\frac 12\sigma_J^2-\mu_J\right)\\
\label{eq:QVjump}
&= (.01\times VIX_n)^2 +\lambda_J\left((\log(1+\mu_J)-.5\sigma_J^2)^2+2\log(1+\mu_J)-2\mu_J\right)\ ,
\end{align}
and so a comparison of VIX to the model's prediction of $QV$ is a meaningful diagnostic. In particular, we should see a slight margin between the two:
\[\frac{1}{\tau^*}\mathbb E^Q[QV_{t_n,\tau^*}|\mathcal F_n^{\Delta t}]-(.01\times VIX_n)^2\]
\begin{equation}
\label{eq:jumpRisk}
 = \lambda_J\left((\log(1+\mu_J)-.5\sigma_J^2)^2+2\log(1+\mu_J)-2\mu_J\right)\ ,
\end{equation}
which is the risk premium placed on jumps. We recalibrate the model everyday, so this risk premium also evolves with time. 

\begin{figure}[htbp]
\begin{minipage}[b]{0.48\linewidth}
\centering
\includegraphics[scale=.6]{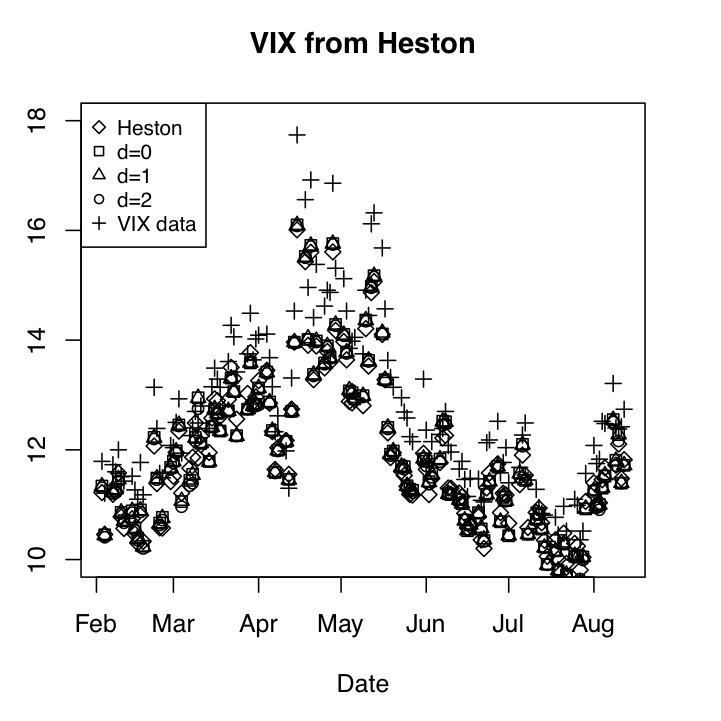}

\end{minipage}
\hspace{0.5cm}
\begin{minipage}[b]{0.45\linewidth}
\centering
\includegraphics[scale=.6]{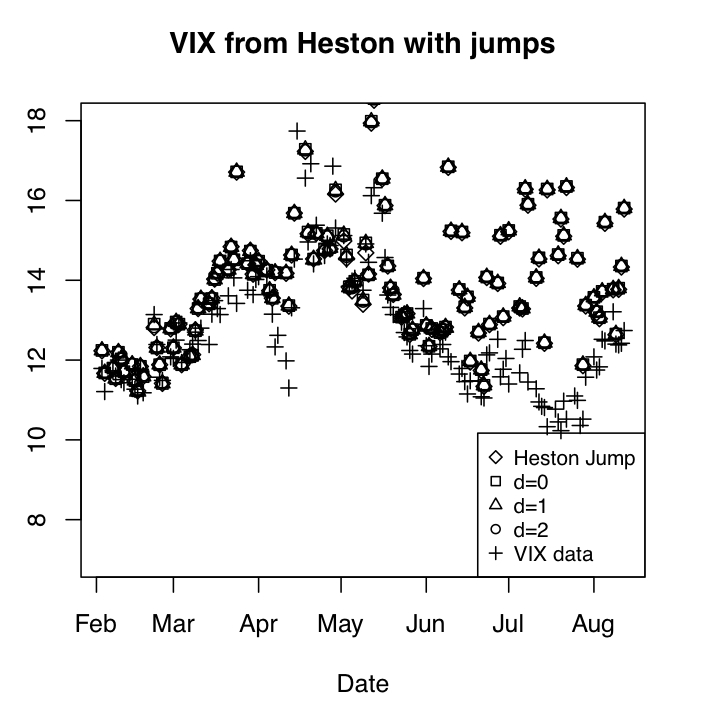}

\end{minipage}
\caption{\small \textbf{Left:} Here we see $VIX$ compared to $100\times$ the square-root of the swap rate under the Heston model, that is, the VIX time series plotted alongside $100\times\sqrt{\bar X+\frac{\mathbb E^Q[X_n|\mathcal F_n^{\Delta t}]-\bar X}{\kappa \tau^*}(1-e^{-\kappa \tau^*})}$ with $\mathbb E^Q[X_n|\mathcal F_n^{\Delta t}]$ approximated using the implied $\widehat\phi_n$ This modelled swap rate has some bias in tracking the VIX, but overall is consistent in it's behavior. \textbf{Right:} Here we see VIX compared to $100\times$ the square-root of the swap rate under the Heston model with jumps as given by equation \eqref{eq:jumpRisk}. The jump model's swap rate is really accurate for the first part of the year, but the two time series separate in the later months. Poor tracking in the later months is due possibly to over fitting, or to an increase in the jump-risk premium that is not present in the VIX.}
\label{fig:hestonVIX}
\end{figure}

In Figure \ref{fig:hestonVIX}, the left-hand plot shows the time series of the VIX alongside the time series of 100$\times$ the square-root of the diffusion component of equation \eqref{eq:proxyQV} using $\widehat\phi_n$ implied by Heston without jumps. In the same figure, the right-hand plot is the same as the left, but uses $\widehat\phi_n$ implied by the Heston model with jumps and includes the jump components of equation\eqref{eq:proxyQV}. The Heston model without jumps appears to be more consistent with the VIX than the model with jumps, particularly in the months of June, July and August. We saw in Figure \ref{fig:hestonMeanStd} that the implied density mean of $X_n$ was similar under both models, and so it must be that the jump model's inconsistencies with VIX are caused by the estimated jump parameters. Empirically, it was observed in \cite{aitSahaliaMancini} and independently in \cite{todorov2010} that there is a significant jump-risk premium in variance swaps that is not accounted for by the VIX. Indeed, from Table \ref{tab:VIXtimeSeries} we see that the Heston model without jumps gives a lower prediction of 30-day variance than the VIX. On the other hand, we see from the table that the Heston model with jumps predicts 30-day variance that is higher than the VIX, which indicates that the fit is picking up the jump-risk premium. However, the erratic behavior in the jump model's variance-swap rate during the later months is confusing and difficult to explain. Hence, while we are fairly certain that the inclusion of the jumps is a good idea and we are able to see the jump-risk premium in Table \ref{tab:VIXtimeSeries}, the right-hand plot in Figure \ref{fig:hestonVIX} raises some concerns (which we address below).
%\begin{table}[!h]
%\caption{\small \textbf{Relative Bias and Error for the Time Series of Modeled VIX}}
%\begin{center}
%\begin{tabular}{c|c|c|c|c}
%\label{tab:VIXtimeSeries}
%&\multicolumn{2}{c|}{Heston}&\multicolumn{2}{c}{Heston with Jumps}\\
%&bias&error&bias&error\\
%\hline
%$x_0$&-0.0594&0.0261&0.1407&0.2107\\
%$d=0$&-0.0601&0.0266&0.1421&0.2109\\
%$d=1$&-0.0602& 0.0265&0.1422& 0.2109\\
%$d=2$&-0.0618& 0.0259&0.1430& 0.2108
%\end{tabular}
%\end{center}
%\caption{\small The relative bias and error for the implied density's computation of VIX. The column labeled `bias' lists the time series mean of $\frac{100\times\sqrt{ \frac{1}{T}\mathbb E^Q[QV_{t_n,T}|\mathcal F_n^{\Delta t}]}- VIX_n}{ VIX_n}$, and the column labeled `error' looks at the time series standard error. Each row represents the implied filter $ \frac{1}{T}\mathbb E^Q[QV_{n,T}|\mathcal F_n^{\Delta t}]$ obtained using varying degree of regularity in $\hat\phi_n$; the row labeled `$x_0$' has the estimates obtained using $\hat\phi_n$ equal to a point-mass.}
%\end{table}
\begin{table}[!h]
\caption{\small \textbf{Variance Swap Relative Bias for the Time Series of Modeled VIX}}
\begin{center}
\begin{tabular}{c|c|c}
\label{tab:VIXtimeSeries}
&\multicolumn{1}{c|}{Heston}&\multicolumn{1}{c}{Heston with Jumps}\\
\hline
$x_0$&-0.0594&0.1407\\
$d=0$&-0.0601&0.1421\\
$d=1$&-0.0602&0.1422\\
$d=2$&-0.0618&0.1430
\end{tabular}
\end{center}
\caption{\small The relative bias for the implied density's computation of VIX. The columns list the time series mean of $\frac{100\times\sqrt{ \frac{1}{\tau^*}\mathbb E^Q[QV_{t_n,\tau^*}|\mathcal F_n^{\Delta t}]}- VIX_n}{ VIX_n}$. Each row represents the proxy of the filter $ \frac{1}{\tau^*}\mathbb E^Q[QV_{n,\tau^*}|\mathcal F_n^{\Delta t}]$ obtained using varying degree of regularity in $\hat\phi_n$; the row labeled `$x_0$' has the estimates obtained using $\hat\phi_n$ equal to a point-mass.}
\end{table}

\begin{figure}[htbp] %  figure placement: here, top, bottom, or page
   \centering
   \includegraphics[scale=.6]{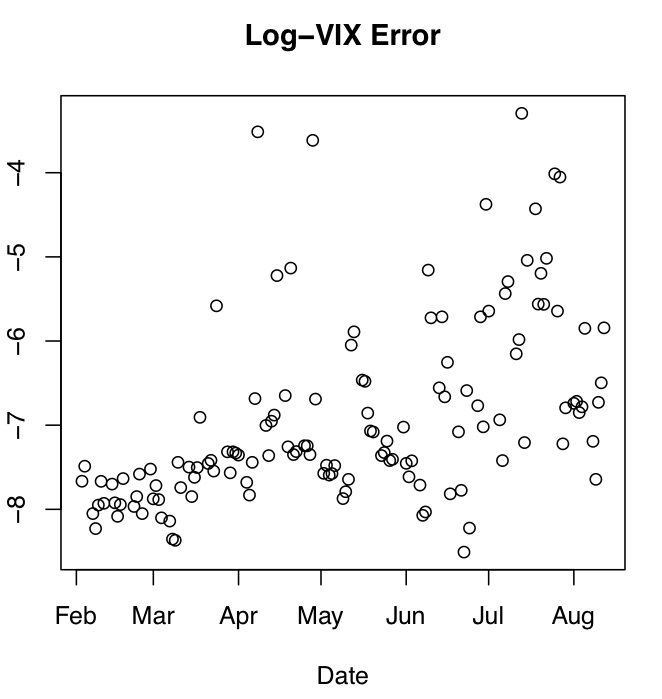} 
   \caption{\small This plot is the logarithm of $\frac{1}{\tau^*}\mathbb E^Q[QV_{t_n,\tau^*}|\mathcal F_n^{\Delta t}]-(.01\times VIX_n)^2$, which is the same thing as the logarithm of jump-risk premium given in equation \eqref{eq:jumpRisk}.}
   \label{fig:RPerr}
\end{figure}

In general, jumps add more degrees of freedom to the model, and so there is expected to be a better in-sample fit to the data (which we indeed saw in the implied volatilities of Figures \ref{fig:ivol} and \ref{fig:ivolJump}). However, additional factors should not be added gratuitously, and should have an interpretation that relates them to the stylized facts observed in the market. In Figure \ref{fig:hestonVIX}, it appears that the Heston model with jumps does a very good job of tracking the VIX during the first half of 2005, but fails to track in the later months. It is possible that the months of June, July and August have been over fit by the model, but it is equally possible that variance swap contracts entered in these months provided investors with insurance against jumps that was not provided by the VIX. Indeed, Figure \ref{fig:RPerr} shows the log-difference between the right-hand side of equation \eqref{eq:jumpRisk}, in which there appears to be an increase in the jump-risk premium in the later months. Nevertheless, further suspicions of over fitting comes by noticing that the jump-intensity estimates in Figure \ref{fig:hestonJumpParamsEst} are decreasing over time, while the jump-risk premium in Figure \ref{fig:RPerr} is increasing; we are suspicious because the model is rich enough that there are likely to be parameterizations wherein the jump intensity increases with the jump premium.
 
In summary, we find that implied $\widehat\phi_n$ from both the Heston and Heston model with jumps have potential to be consistent with the VIX. The results indicate that the more parsimonious Heston model without jumps consistently underprices  the variance swap, which is seen as we compare the two time series' of the implied density's swap rate and the VIX. When jumps are included, the variance-swap rate consistently adds a jump-risk premium during the first half of 2005, but this risk premium becomes erratic in the later months. We suspect that there has been over fitting to the data by the jump-model, but we do not contend that the jump-model is mis-specified; we simply think that the jump model during this time period has exhibited some unstable behavior that warrants further investigation.

%%%%%%%%%%%%%%%%%%%%%%%%%%%%%%%%%%%%%%%%%%%%%%%%%%%%%%%%%%%%%%%%%%%%%%%%%%%%%%%%%%%%%%%%%%
%%%%%%%%%%%%%%%%%%%%%%%%%%%%%%%%%%%%%%%%%%%%%%%%%%%%%%%%%%%%%%%%%%%%%%%%%%%%%%%%%%%%%%%%%%
%%%%%%%%%%%%%%%%%%%%%%%%%%%%%%%%%%%%%%%%%%%%%%%%%%%%%%%%%%%%%%%%%%%%%%%%%%%%%%%%%%%%%%%%%%

\section{Summary \& Conclusions}
We have used a Tykhonov regularization to invert the risk-neutral prices  of derivatives, and have obtained an implied filtering density on the hidden state of volatility. The method was shown to be effective in simulations where the measurements are precise and parameters are known. The solution was also shown to be  accurate in estimating the 1st and 2nd filtering moments in the presence of measurement imprecision and parameter uncertainty. 

When applied to SPX data, it appears that the model's fitted parameters and the implied density can pick up the maturity cycles in the options data, which we interpret as a volatility-uncertainty premium. Also, the fitted model parameters and the implied density's 1st and 2nd moments for the Heston model, both with and without jumps, produce a modeled rate for 30 day variance swaps on SPX that exhibited some consistency with the VIX. Overall, we find the Heston model without jumps to be straight forward to interpret because the implied objects that are easy to interpret. In contrast, the Heston model with jumps conceals the maturity cycles and implies parameters that have a less-simple explanation, and hence, it is possible that the jump model has over fit the data.

In the future, it would be interesting to explore different types of data, such as the term structures on variance swaps or commodity futures. It would also be interesting to explore richer models, and perhaps find a way to quickly solve the inverse problem using fast algorithms. An important issue that  should be addressed is the challenge of obtaining a hedging portfolio when volatility is not observed.

%%%%%%%%%%%%%%%%%%%%%%%%%%%%%%%%%%%%%%%%%%%%%%%%%%%%%%%%%%%%%%%%%%%%%%%%%%%%%%%%%%%%%%%%%%%%%%%%%%%%%%%
%%%%%%%%%%%%%%%%%%%%%%%%%%%%%%%%%%%%%%%%%%%%%%%%%%%%%%%%%%%%%%%%%%%%%%%%%%%%%%%%%%%%%%%%%%%%%%%%%%%%%%%
\newpage

\begin{table}
\center
\caption{\small \textbf{Estimated Parameters for Heston Model without Jumps}}
\begin{tabular}{c|c|c|c|c}
\label{tab:hestonParams}
date&$\kappa$&$\bar X$&$\gamma$&$\rho$\\
\hline
2/3 & 5.56 & 0.02 & 0.45 & -0.66 \\
2/4 & 6.33 & 0.02 & 0.47 & -0.68 \\
2/7 & 8.21 & 0.02 & 0.53 & -0.67 \\
2/8 & 10.47 & 0.02 & 0.57 & -0.65 \\
2/9 & 12.01 & 0.02 & 0.62 & -0.66 \\
2/10 & 4.27 & 0.02 & 0.39 & -0.69 \\
2/11 & 4.7 & 0.02 & 0.4 & -0.68 \\
2/14 & 4.51 & 0.02 & 0.4 & -0.7 \\
2/15 & 5.61 & 0.02 & 0.43 & -0.68 \\
2/16 & 6.01 & 0.02 & 0.44 & -0.68 \\
2/17 & 5.88 & 0.02 & 0.45 & -0.66 \\
2/18 & 6.37 & 0.02 & 0.47 & -0.66 \\
2/22 & 7.26 & 0.02 & 0.51 & -0.67 \\
2/23 & 6.27 & 0.02 & 0.49 & -0.65 \\
2/24 & 5.75 & 0.02 & 0.46 & -0.64 \\
2/25 & 5.87 & 0.02 & 0.46 & -0.68 \\
2/28 & 5.68 & 0.02 & 0.46 & -0.69 \\
3/1 & 5.67 & 0.02 & 0.46 & -0.65 \\
3/2 & 9.63 & 0.02 & 0.59 & -0.65 \\
3/3 & 10.09 & 0.02 & 0.6 & -0.61 \\
3/4 & 14.58 & 0.02 & 0.7 & -0.7 \\
3/7 & 17.91 & 0.02 & 0.77 & -0.67 \\
3/8 & 24.58 & 0.02 & 0.9 & -0.67 \\
3/9 & 25.59 & 0.02 & 0.93 & -0.61 \\
3/10 & 12.51 & 0.02 & 0.66 & -0.62 \\
3/11 & 4.81 & 0.02 & 0.44 & -0.59 \\
3/14 & 5.58 & 0.02 & 0.47 & -0.61

\end{tabular}
\end{table}

\begin{table}
\center
\caption{\small \textbf{Estimated Parameters for Heston Model with Jumps (Equation \eqref{eq:hestonModel})}}
\begin{tabular}{c|c|c|c|c|c|c|c}
\label{tab:hestonJumpParams}
date&$\kappa$&$\bar X$&$\gamma$&$\rho$&$\lambda_J$&$\mu_J$&$\sigma_J$\\
\hline
2/3 & 1.92 & 0.02 & 0.26 & -0.67 & 0.07 & -0.1 & 0.23 \\
2/4 & 2.34 & 0.02 & 0.29 & -0.65 & 0.05 & -0.13 & 0.27 \\
2/7 & 1.38 & 0.02 & 0.23 & -0.68 & 0.11 & -0.08 & 0.17 \\
2/8 & 1.42 & 0.02 & 0.22 & -0.69 & 0.12 & -0.07 & 0.16 \\
2/9 & 1.93 & 0.02 & 0.25 & -0.71 & 0.08 & -0.08 & 0.21 \\
2/10 & 1.14 & 0.02 & 0.2 & -0.65 & 0.08 & -0.11 & 0.2 \\
2/11 & 0.94 & 0.02 & 0.18 & -0.68 & 0.12 & -0.08 & 0.18 \\
2/14 & 1.21 & 0.02 & 0.2 & -0.64 & 0.09 & -0.11 & 0.2 \\
2/15 & 1.18 & 0.02 & 0.2 & -0.66 & 0.11 & -0.09 & 0.18 \\
2/16 & 1.27 & 0.02 & 0.21 & -0.64 & 0.11 & -0.08 & 0.17 \\
2/17 & 0.71 & 0.02 & 0.17 & -0.65 & 0.15 & -0.08 & 0.16 \\
2/18 & 3.17 & 0.02 & 0.31 & -0.61 & 0.06 & -0.12 & 0.22 \\
2/22 & 1.08 & 0.02 & 0.19 & -0.71 & 0.26 & -0.06 & 0.14 \\
2/23 & 1.66 & 0.02 & 0.24 & -0.59 & 0.16 & -0.09 & 0.15 \\
2/24 & 2.87 & 0.02 & 0.31 & -0.6 & 0.06 & -0.12 & 0.24 \\
2/25 & 1.94 & 0.02 & 0.26 & -0.62 & 0.08 & -0.09 & 0.19 \\
2/28 & 0.85 & 0.02 & 0.17 & -0.61 & 0.17 & -0.1 & 0.16 \\
3/1 & 1.48 & 0.02 & 0.23 & -0.6 & 0.11 & -0.09 & 0.17 \\
3/2 & 2.2 & 0.02 & 0.27 & -0.6 & 0.09 & -0.09 & 0.21 \\
3/3 & 1.36 & 0.02 & 0.22 & -0.6 & 0.15 & -0.08 & 0.17 \\
3/4 & 2.88 & 0.02 & 0.3 & -0.58 & 0.17 & -0.07 & 0.15 \\
3/7 & 1.73 & 0.02 & 0.24 & -0.57 & 0.22 & -0.07 & 0.13 \\
3/8 & 1.46 & 0.02 & 0.22 & -0.62 & 0.26 & -0.05 & 0.12 \\
3/9 & 1.49 & 0.02 & 0.22 & -0.64 & 0.33 & -0.05 & 0.11 \\
3/10 & 2.32 & 0.02 & 0.25 & -0.58 & 0.08 & -0.11 & 0.23 \\
3/11 & 0.7 & 0.02 & 0.16 & -0.66 & 0.19 & -0.07 & 0.17 \\
3/14 & 2.28 & 0.02 & 0.25 & -0.61 & 0.09 & -0.1 & 0.22 
\end{tabular}
\end{table}

\small
\bibliography{myRefs.bib} 

\end{document}